\documentclass[]{pasj01}
\usepackage{enumerate}
\usepackage{url}
\usepackage{color}
\usepackage{lineno}
\Received{$\langle$reception date$\rangle$}
\Accepted{$\langle$acception date$\rangle$}
\Published{$\langle$publication date$\rangle$}
\begin{document}

%\title{Shock heating of an early merging cluster,CIZA J1358.9 -4750}
\title{XMM-Newton view of the shock heating in an early merging cluster, CIZA~J1358.9$-$4750}
%\author{Yuki\,OMIYA,\,Kazuhiro\,NAKAZAWA,\,Kyoko\,Matsusita,\\Shogo\,KOBAYASHI,\,Nobuhiro\,OKABE,\,Kosuke\,SATO,\\Takayuki\,TAMURA,\,Yutaka\,FUJITA\,and\,Liyi\,Gu\,Tetsu\,KITAYAMA\\Takuya\,AKAHORI\,and\,Kohei\,KURAHARA\,and\,Tomohiro\,YAMAGUCHI}%
\author{ Yuki \textsc{Omiya},\altaffilmark{1,}$^{*}$
 Kazuhiro \textsc{Nakazawa},\altaffilmark{1,10} 
 Kyoko \textsc{Matsushita},\altaffilmark{2}
 Shogo \textsc{B. Kobayashi},\altaffilmark{2}
 Nobuhiro \textsc{Okabe},\altaffilmark{3}
 Kosuke \textsc{Sato},\altaffilmark{4}
 Takayuki \textsc{Tamura},\altaffilmark{5}
 Yutaka \textsc{Fujita},\altaffilmark{6}
 Liyi \textsc{Gu},\altaffilmark{7}
 Tetsu \textsc{Kitayama},\altaffilmark{8}
 Takuya \textsc{Akahori},\altaffilmark{9}
 Kohei \textsc{Kurahara},\altaffilmark{9} and
 Tomohiro \textsc{Yamaguchi}\altaffilmark{1} }
\altaffiltext{1}{Departure of Physics, Nagoya University, Furo-cho, Chikusa-ku, Nagoya, Aichi 464-8601, Japan}
\altaffiltext{2}{Department of Physics, Tokyo University of Science, 1-3 Kagurazaka, Shinjuku-ku, Tokyo 162-8601, Japan}
\altaffiltext{3}{Department of Physics, Hiroshima University, 1-3-1 Kagamiyama, HigashiHiroshima, Hiroshima 739-8526, Japan}
\altaffiltext{4}{Department of Physics, Saitama University, 255 Shimo-Okubo, Sakura-ku, Saitama 338-8570, Japan}
\altaffiltext{5}{Institute of Space and Astronautical Science, Japan Aerospace Exploration Agency, 3-1-1 Yoshinodai,
Chuo-ku, Sagamihara, Kanagawa 229-8510, Japan}
\altaffiltext{6}{Department of Physics, Graduate School of Science, Tokyo Metropolitan University, 1-1 Minami-Osawa, Hachioji-shi, Tokyo 192-0397, Japan}
\altaffiltext{7}{SRON Netherlands Institute for Space Research, Utrecht, The Netherlands}
\altaffiltext{8}{Department of Physics, Toho University, 2-2-1 Miyama, Funabashi, Chiba 274-8510, Japan}
\altaffiltext{9}{Mizusawa VLBI Observatory, National Astronomical Observatory of Japan, 2-21-1, Osawa, Mitaka, Tokyo 181-8588, Japan}
\altaffiltext{10}{Kobayashi-Maskawa Institute for the Origin of Particles and the Universe, Furo-cho, Chikusa-ku, Nagoya, Aichi 464-8601, Japan}

\email{omiya$_{-}$y@u.phys.nagoya-u.ac.jp}


\maketitle

\begin{abstract}
%%Merging clusters of galaxies convert large amounts of gravitational energy into 
%gas heating 
%%heating of their intra-cluster matter (ICM) and even non-thermal energy. However, how much of this energy is distributed to each component is not known. %CIZA~J1358.9$-$4750 is located very close (z=0.07) and is the major merger nearby on 
%足す
%the
%sky plane.

%Merging clusters of galaxies convert large amounts of gravitational energy into heating of their intra-cluster matter (ICM) and even non-thermal energy. However, how much of this energy is distributed to each component is not known.
%CIZA~J1358.9$-$4750 
%(CIZA J1359) 
%is a major merger cluster on the sky plane located very close (z=0.07).
%The southeast and northwest clusters are separated by ~1.4 Mpc in the sky, and Suzaku observations showed that there is a candidate shock wave of Mach 1.4 in the middle of the bridging structure between the two clusters (Kato+ 2016). 
%足す
CIZA~J1358.9$-$4750 is a nearby galaxy cluster in the early phase of a major merger.
The two-dimensional temperature map using XMM-{\it Newton} EPIC-PN observation confirms the existence of a high temperature region, which we call the ``hot region'', in the ``bridge region'' connecting the two clusters. 
The $\sim$500 kpc wide region between the southeast and northwest boundaries 
\color{black}
also has higher pseudo pressure compared to the unshocked regions, suggesting the existence of two shocks. 
The southern shock front is clearly visible in the X-ray surface brightness image and has already been reported by \citet{2015PASJ...67...71K}. The northern one, on the other hand, is newly discovered. To evaluate their Mach number, we constructed a three-dimensional toy merger model with overlapping shocked and unshocked components in line of sight. 
%The unshocked ICM conditions were estimated based on those outside the interacting bridge region assuming point symmetry.
%Based on the enhancement ratio of the X-ray normalization, evaluated by two-temperature thermal components fitting to the spectra, coupled with the Rankine-Hugoniot equation, we estimated the shocked ICM parameters, including the densities of shocked and un-shocked regions. 
The unshocked and preshock ICM conditions are estimated based on those outside the interacting bridge region assuming point symmetry.
The hot region spectra are modeled with two-temperature thermal components,
\color{black}
assuming that the shocked condition follows the Rankin-Hugoniot relation with the preshock condition.
%We modeled the hot region spectra with two-temperature thermal components, assuming that the postshock hot region in addition to the unshocked component. Rankine--Hugoniot equation is assumed between the preshock and shocked region, as well.
%We modeled the hot region spectra with two-temperature thermal components, assuming that the enhancement ratio of the X-ray surface brightness can be explained by the postshock hot region in addition to the unshocked component. Rankine--Hugoniot equation is assumed between the preshock and shocked region, as well.
As a result, the shocked region is estimated to have
a line--of--sight depth
\color{black}
of $\sim$ 1 Mpc with a Mach number of $\sim$1.3 in the southeast shock and $\sim$ 1.7 in the northwest shock.
\color{black}
The age of the shock waves is estimated to be $\sim$260 Myr. This three dimensional merger model is consistent with the Sunyaev-Zel'dovich signal obtained using the Planck observation within the CMB fluctuations. The total flow of the kinetic energy of the ICM through the southeast shock was estimated to be $\sim2.2\,\times\,10^{42}~\rm{erg~s^{-1}}$.
Assuming that 10 \% of this energy is converted into ICM turbulence, the line--of--sight velocity dispersion is calculated to be $\sim200~\rm{km~s^{-1}}$,
which is basically resolvable via coming high spectral resolution observations.
\color{black}
%281/300文字
\color{black}
\end{abstract}
%\linenumbers

\section{Introduction}

%In cosmological large-scale structure 
% 今回こちらにコメントしてみました。トラックチェンジをした方が良いかも。
%formation, 
%Galaxy clusters grow through accumulating ambient matters and merging with other clusters.
Galaxy clusters are the largest self-gravitating system in the universe with a mass of $10^{14-15}M_{\odot}$, consisting of hundreds to thousands of galaxies. Nearly 80\% of the mass is made up of dark matter, with member galaxies making up only $\sim$ 3\% (e.g., \cite{2005ApJ...628..655V}; \cite{2009ApJ...693.1142S}; \cite{2013MNRAS.429.3288S}; \cite{2022PASJ...74..175A}). Most the hadronic matter is in the form of high temperature plasma, called an intracluster medium (ICM). Galaxy clusters are not isolated systems,  but rather grow as they 
%take in 
accumulate
matter from their surroundings through gravitational interactions. 
One of the most common phenomena is a ``major merger'', in which two galaxy clusters with close masses collide almost head-on.
%Among these phenomena,
%"major merger"
%``major merger''
%is a phenomenon in which two %clusters with similar mass %collides
%collide nearly face to face.
Depending on the mass of the cluster, a major merger can release up to $10^{64}$ erg of gravitational energy (e.g., \cite{2004cosp...35.3617R}; \cite{2009MNRAS.399..410P}; \cite{2012MNRAS.420.2120M}).

Numerical simulations have shown that a major merger event generates two types of shock waves (\cite{2000ApJ...542..608M}; \cite{2003ApJ...593..599R}; \cite{2008ApJ...689.1063S}). 
When two clusters merge facing each other, the ICM 
in between
is adiabatically compressed.
%and when
%When
As
the collision velocity exceeds the ICM sound speed, two shock fronts are formed (e.g., \cite{2008ApJ...687..951T}; \cite{2008PASJ...60L..19A}; \cite{2010PASJ...62..335A}; \cite{2018ApJ...857...26H}). 
They move forward and backward along the merger axis, and eventually penetrate both systems, 
as seen in Abell 3667 (\cite{2009PASJ...61..339N}; \cite{2010HEAD...11.3424S}; \cite{2012PASJ...64...49A}; \cite{2016arXiv160607433S}; \cite{2022A&A...659A.146D}), Abell 3376 (\cite{2006Sci...314..791B}; \cite{2009PASJ...61S.377K}; \cite{2012PASJ...64...67A}), and CIZA J2242.8+5301 (\cite{2011MmSAI..82..569V}; \cite{2015PASJ...67..114O}). 
\color{black}
Another shock wave, called the ``equatorial shock wave'', appears perpendicular to the merger axis (e.g., \cite{2019PASJ...71...79O}), 
as observed in the merger of 1E 2216.0-0401 and 1E 2215.7-0404 (\cite{2019NatAs...3..838G}).
\color{black}
%(e.g. 1E 2216.0-0401 and 1E 2215.7-0404 merger,  \cite{2019NatAs...3..838G}). 
%hese 
These
shocks cause ICM heating, as well as ICM turbulence excitation, particle acceleration, and magnetic field amplification (
e.g.,
\cite{2007PhR...443....1M}; \cite{2010Sci...330..347V}).
%. Athough 
However,
the branching energy ratio of these four phenomena is not well understood. 

%The most effective way 
In order to estimate 
the
energy input,
%is to 
it is necessary to
construct the 
dynamical structures 
of merging clusters.
Among them, 
% merge という単語が多すぎる
%merging clusters
those in the early phase are relatively simple to model. 
%In the early stages of a merger, the shock wave has not yet spread to the outer edges of the cluster, so the properties of the cluster before the merger 
%are 
In the early phase of 
a major merger, 
\color{black}
shock waves have not yet reached the outer edges of the cluster, and the outer regions mostly
retain 
\color{black}
the pre-merger condition.
\color{black}
Therefore, the properties of the ``pre-merger'' cluster
can be 
well inferred. 
%Since 
Once the initial conditions are known, it is relatively easy to elucidate the structure of the cluster in merging. 
%On the other hand, 
When
%the shock waves are characterized by young shocks and
a shock wave is young,
%and 
%has small Mach numbers. 
its Mach number is typically limited to less than 2 (\cite{2010PASJ...62..335A}).
In the late phase, it reaches $\sim$4 (\cite{2001ApJ...561..621R}).
\color{black}
%不要かな。。。微妙、あとで考えましょう。
%The small indeterminacy of the merger model makes it possible to study in detail the shock waves that have not been generated for a long time.

%In particular, clarifying the structure of merging galaxy clusters in their early stages, which shock waves have not yet spread to the outer, is expected to elucidate the distribution of thermal and non-thermal energy.

%その中でも衝突初期段階の銀河団は比較的モデル化しやすい。衝突の初期段階は衝撃波が銀河団外縁部まで広がっていないため衝突前の銀河団の状態がよくわかる。初期値がわかっているため衝突後の構造を推測しやすい。一方で初期の衝突銀河団の衝撃波はショックが若く、マッハ数が小さいという特徴がある。そのような生成されて時間がたっていない衝撃波についてはmerger modelの不定性の小ささと合わせて詳しく調べることが可能になる。

%CIZA J1358.9 -4750 
CIZA J1358.9$-4750$ 
(hereafter, CIZA J1359)
%is one of the brightest cluster
is a nearby galaxy cluster (z=0.0074; \cite{2011A&A...534A.109P})
%[hereafter, CIZA J1359 (z=0.0074; \cite{Piffaretti2011})] is one of nearby brightest cluster
%
listed
in the Zone of Avoidance (CIZA) catalog 
(\cite{2002ApJ...580..774E}, \cite{2007ApJ...662..224K}).
%, in the phase of the major merger.
% 
\color{black}
%CIZAカタログ(Ebeling et al. 2002; Kocevski et al. 2007)
The
%center
X-ray centroid of the southeastern (SE) cluster 
is 14' ($\sim$ 1.2 Mpc) away from the northwestern (NW) cluster on the sky plane.
%is located $14'$ (or 1.2 Mpc) apart in the sky plane from the northwest one. 
The bridge region between the two is clearly visible in the X-ray image in figure \ref{fig:fig1}.
%Also, a bridge region between them is clearly visible in the X-ray image shown in figure 1.
%図示
%The two galaxies (WISEA J135903.81-475131.1 and WISEA J135810.81-474124.7) at the center of southeast and northwest clusters have optical redshift of 0.0745 and 0.0709, respectively (\cite{D. Heath Jones2009})
%, so the two clusters are merging nearby on sky plane. 
%（ここは最新の解析に合わせて文言を変えましょう。案１：このままKato et al の結論を述べた後に、新たな段落で赤方偏移のまとめをして、「やはりsky plane 上」にあると考えて良いというか、あるいは
%案２：この段落でまず、Kato et al の結論を示した上で、「なぜなら＊＊、＊＊」という書き方にして、その後半で redshift の話をするとか）

%Suzaku observation 
\citet{2015PASJ...67...71K} observed CIZA J1359 with the Suzaku X-ray observatory and 
found 
a ``brightness jump''
in the southeast of
the
bridge region. %has X-ray 
%%shown
%has 
%a clear ``line'' of X-ray enhancement 
%and 
%It associated with 
%high 
%higher temperatures than the neighboring regions
%using Suzaku observation. 
%It is identified the front shock and derived a Mach 
%and 
%from the temperature jump
%a Mach 
%number 
This ``brightness jump'' was associated with a temperature jump, and the Mach number was calculated to be
1.32~$\pm$~0.22.
\color{black}
%
%was suggested. 
%Thus,
Based on its typical ``two clusters and a bridge'' morphology and the presence of the narrow enhancement region, CIZA J1359 is claimed to be an early merger with its merger axis located approximately along the sky plane. 
%Because of its prototypical ``two-clusters and bridge'' morphology and existence of narrow enhanced region, 
%CIZA J1359 is 
%concluded an early merging cluster 
%claimed to be an
%prototypical 
%early merger,
%with their merging axis located nearly along the sky plane  .
%(\cite{Kato2015}). 
%後退速度と衝突速度からsky planeからの傾きを求めると45度程度-->傾いていないと仮定して3Dmodelingしている
\citet{2018PASJ...70...53A} explored the existence of %radio halos, mini-halos, and relics 
extended radio emission
using the 16~cm band Australia Telescope Compact Array.
Their main objective was to detect possible relic radio emissions, but they did not find any relic, halo, or mini-halo.
Similarly, \citet{2022MNRAS.514.5969K} reported that there was no detection of diffuse cluster radio emission in the bridge region.
The results suggest that CIZA J1359 
is still 
\color{black}
in the early phase of the merger and that the shock front has not yet reached 
the outer edges of the cluster. 
\color{black}

In this paper, we report the results of our analysis based on a deep XMM--{\it Newton} observation of this cluster.
The outline of this %pape
paper is as %follow.
follows.
In Section 2, we describe the construction of a two-dimensional (2D) thermodynamic map
%%add
%of temperature and surface brightness,
and the discovery of the presence of 
two shock waves $\sim 500$~kpc apart.
\color{black}
In Section 3, we construct a three-dimensional (3D) model and discuss shocked conditions at two shock fronts. 
We assume $H_{0}~=~70~\rm{km~s^{-1}~Mpc^{-1}}$,$\Omega_{M}~=~0.3$ and $\Omega_{\Lambda}~=~0.7$ (1~arcmin~=~80.34~kpc at z~=~0.0074) throughout this paper. Errors are given at the 68\% confidence ($1\sigma$) level unless otherwise stated.
%2次元の熱力学マップから得られる2つの衝撃波を示す。
%3次元モデルを構築し、post-shockコンディションを状態を議論する

%post shock condition region pre shock condition regionを緑色のラインで示す。2.3章

\section{Data analysis}

\subsection{XMM-{\it Newton} observation}

In this work, we used a 99~ks XMM-{\it Newton} observation 
%CIZA J1359 was observed for 99~ks by the %$\it{XMM-Newton}$ 
%XMM-{\it Newton} 
%satellite 
on August 24$-$25, 2016, pointing to the 
%cluster center 
center of the bridge region (obsid:0784980101). 
Among the European Photon Imaging Camera (EPIC)
%in the XMM, ,
onboard XMM, EPIC-MOS1 CCD3 and CCD6, which cover the bridge region, were lost to the micrometeorite hit (\cite{2006ESASP.604..943A}; \cite{2008A&A...478..615S}) and EPIC-MOS2 CCD4 and CCD5 are possibly in an anomalous state (e.g., \cite{2012A&A...537A..39S}; \cite{2008A&A...478..575K}). 
The effective area of EPIC-PN is larger than that of the EPIC-MOS. %, thus covering the entire CIZAJ1359.
In addition, there is a calibration inconsistency between EPIC-MOS and PN on the hardness of the spectra, especially in the higher energy band (e.g., \cite{2017AJ....153....2M}).
MOS tends to give a systemically harder photon index than PN and the Suzaku XIS detectors in this energy band. In ICM thermal fitting, the effect gives a much higher temperature than those obtained by Suzaku, especially when the ICM temperature is high (\cite{2015PASJ...67...71K}, \cite{KatoD}). 
To simplify this issue and because the EPIC-PN camera provides nearly two-thirds of the photon statistics in the bridge region of this observation, in this work, we focus on EPIC-PN data.
%Also, the previous works reported the analysis results of three EPIC instruments are inconsistent with each other (e.g. \cite{2011A&A...525A..25T}; \cite{2017AJ....153....2M}; \cite{2020A&A...642A..89Z}).
%Based on these facts, in this paper, we concentrate on EPIC-PN data combining with
%Combined with 
%its high quantum efficiency (\cite{2001A&A...365L..18S}).
\color{black}
%in this paper we concentrate on EPIC-PN data.
%%we have only used the EPIC-pn data for its high detection efficiency.
%
%mos1は欠陥チップが存在
%加藤D論ではpnは北側にもエンハンスがあるので
%We used the observation of PN, which is the back-illuminated CCD of the Europian Photon Imaging Camera (EPIC) instruments.

\begin{figure}[tp]
 \begin{center}
 %%
 %%
 %\centering
  \includegraphics[width=8cm]{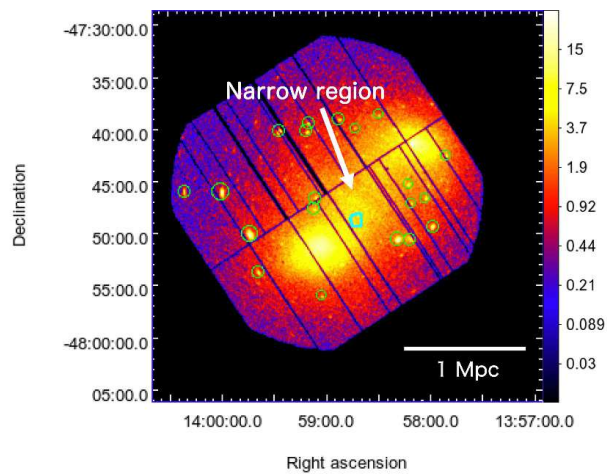}
 \end{center}
 \caption{A total counts X-ray image\color{black} of CIZA J1359 observed with XMM--{\it Newton} EPIC--PN detector in 0.5 -- 12 keV, smoothed with a Gaussian kernel with $\sigma$ = 2.5". The colorbar have the unit of counts. The green circles indicate the point source, which we removed in the analysis. The cyan polygon is the region from which the spectrum in figure \ref{fig:spectrum} was extracted, corresponding to the SE3 in the upper right panel of figure \ref{fig:fig2}.}
 \label{fig:fig1}
%スペクトルを示す領域を付け加える
\end{figure}

Standard processing of the observed data was performed using the Extended Source Analysis Software 
(ESAS), version 18.0.0, 
\color{black}
as described by \citet{2008A&A...478..615S}, following the ``cookbook for analysis procedures for XMM-Newton EPIC'' \footnote{\url{http://heasarc.gsfc.nasa.gov/docs/xmm/esas/cookbook}}.
%The observational data set was reduced and analysed using the Extended Source Analysis Software (v18.0.0) as described in \cite{2008A&A...478..615S}, 
%
Calibrated photon event files were produced using the {\it epchain} task.
We used the {\it pn-filter} task to remove proton contamination and produced a 65~ks Good Time Interval file.
%The image shown in figure~\ref{fig:fig1} is created the {\it pn-spectra} task
Figure \ref{fig:fig1} is a total counts image in 0.5 -- 12 keV smoothed with $\sigma$=2.5" Gaussian kernel created using the {\it pn-spectra} task.
%The {\it pn-spectra} task was created a 0.5--12.0 keV image shown in figure~\ref{fig:fig1}, which is smoothed with $\sigma$=2.5 arcsec Gaussian kernel.
%The spectrum extracted from the blue region in figure ~\ref{fig:fig1} is shown in  figure~\ref{fig:fig1} .
\color{black}
%We ran {\it epchain} and {\it pn-filter} to remove
%
%periods of high soft-proton flux by building 20 clipping Good Time Interval %(GTI) 
%files in the hard band (12-14 keV). 
%We then created
%
%an
%image 
%which extraction was restricted to the 500$-$1000 keV band.
%using the 0.5--12.0 keV band. 
%The obtained image, 
%After smoothed with $\sigma$=2.5 arcsec Gaussian kernel,
%
%the image
%is shown in figure~\ref{fig:fig1}.
%The blue circles indicate the point sources detected by the CHEESE task with fluxes above $10^{-14} \rm{erg~cm^{-2 }~s^{-1}}$ and with PSF model counts extended to 1/4 of the background.

\begin{figure}[tp]
 \begin{center}
  \includegraphics[width=8cm]{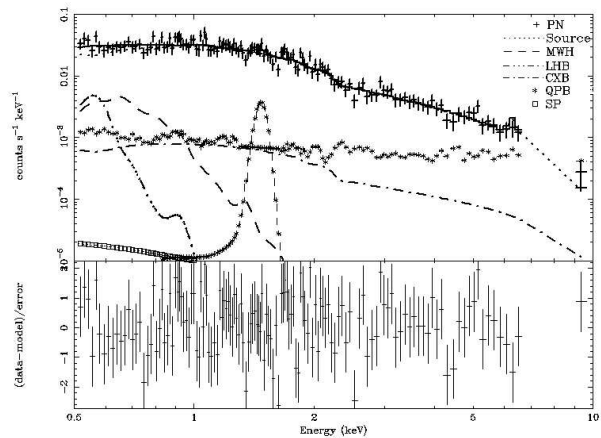}
 \end{center}
 \caption{The EPIC-PN spectra with the QPB component subtracted in the polygon region indicated by cyan in figure \ref{fig:fig1}. The solid line represents the overall best-fit model. Its individual modeled components (source and backgrounds) are shown in the legend.}
 \label{fig:spectrum}
%スペクトルを示す領域を付け加える
\end{figure}

\begin{figure*}[t]
  \begin{center}
  \includegraphics[width=17cm]{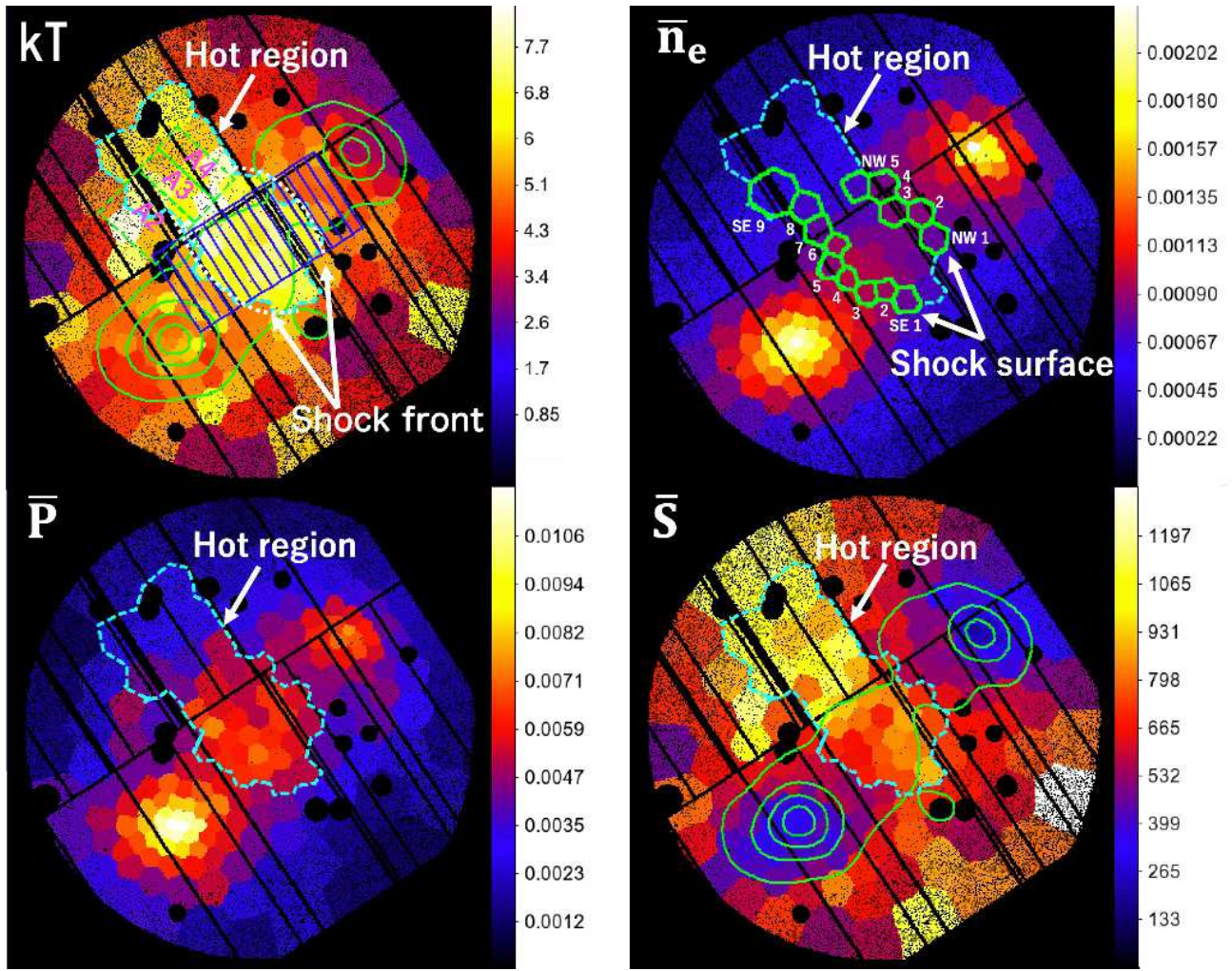}
  \end{center}
	%\begin{minipage}{0.5\hsize}
	%\begin{center} 
	
		%\includegraphics[width=0.95\columnwidth]{paramap_temp-radec_ver.}
	%\end{center}
	%\end{minipage}
	%\begin{minipage}{0.5\hsize}
	%\begin{center} 
		%\includegraphics[width=0.95\columnwidth]{paramap_electden-radec_ver.png}
	%\end{center}
	%\end{minipage}
	%\hfill
	%\begin{minipage}{0.5\hsize}
	%begin{center} 
	%	\includegraphics[width=0.95\columnwidth]{paramap_pressure-radec_ver.png}
	%\end{center}
	%\end{minipage}
	%\begin{minipage}{0.5\hsize}
	%\begin{center} 
	%	\includegraphics[width=0.95\columnwidth]{paramap_entropy-radec_ver.png}
	%\end{center}
	%\end{minipage}
\caption{Thermodynamic maps with regions divided by the WVT binning algorithm. The upper-left panel shows a temperature map overlaid with green contours for X-ray surface brightness. The cyan dotted line shows the hot region we define, and the blue boxes are the projection regions discussed in Section \ref{stimating the mach number of two-shocks}. Green-dotted boxes indicate the regions of A2, A3, and A4 given in \citet{2015PASJ...67...71K}, respectively. The pseudo-density (upper right), pseudo-pressure (lower left), and pseudo-entropy (lower right) maps assume that the ICM is uniformly spread over the line-of-sight depth of 1 Mpc. Units of 2D temperature, pseudo-density, pseudo-pressure, and pseudo-entropy maps are keV,  $\textrm{cm}^{-3}$ $\times$ $(l/1~\textrm{Mpc})^{-1/2}$,  $\textrm{keVcm}^{-3}$ $\times$ $(l/1~\textrm{Mpc})^{-1/2}$ and $\textrm{keVcm}^{2}$ $\times$ $(l/1~\textrm{Mpc})^{1/3}$, respectively.}
%$\textrm{cm^{-3}}$ $\times$ $\textrm{(l/1~Mpc)^{-1/2}}$,  $\textrm{keVcm^{-3}}$ $\times$ $\textrm{(l/1~Mpc)^{-1/2}}$ and $\textrm{keVcm^{2}}$ $\times$ $\textrm{(l/1~Mpc)^{1/3}}$
\label{fig:fig2}
%Thermodynamic_mapping
\end{figure*}

\subsection{2D thermodynamic mapping}
\label{section Thermodynamic mapping}

To characterize the ICM in the bridge region, 
we generated
thermodynamic maps.
%ping was performed using the Weighted Voronoi Tesselations (WVT) binning algorithm (\cite{2006MNRAS.368..497D}). This algorithm divides the field of view into smaller regions with close signal-to-noise ratios (S/N). 
%To explore the properties of the ICM in the bridge region, we conducted a thermodynamic mapping adapted Weighted Voronoi Tesselations (WVT) binning algorithm (\cite{2006MNRAS.368..497D}), which divides the field of view into small regions with a similar signal-to-noise (S/N) ratio. 
%A circular region with 
We first selected a circular region with a radius of 12' ($\sim$1 Mpc) centered on the bridge region, $(\alpha, \delta)_{\rm J2000.0} = ($\timeform{13h58m39s.3}, $-$\timeform{47D46'29''.1}), 
%was taken and 
%divided so that each region had about 2500 counts
and divided it into about 180 regions, each containing approximately 2500 counts, 
based on the Weighted Voronoi Tesselations (WVT) binning algorithm (\cite{2006MNRAS.368..497D}).
The spectra of each subregion were extracted
%from 
%the 
%each of the 
%individual
%sub regions 
and 
%fit it with radiation model (source+background) using xspec v12.12.0.
fitted with a thermal emission model,
using xspec v12.12.0 (\cite{1996ASPC..101...17A}).

As an example of spectral fitting, figure \ref{fig:spectrum} shows the results in the region indicated by cyan in figure \ref{fig:fig1}.
\color{black}
Quiescent particle background (QPB) spectra were generated by the {\it pn\_back} task and subtracted from the source spectra.
The Al K$\alpha$ emission line was modeled as a Gaussian, and the 7--9 keV band was ignored in the fitting to 
avoid 
\color{black}
other instrumental background lines.
\color{black}
As a sky background, three components were modeled;
Local Hot Bubble (LHB) radiation with a temperature 
%of
fixed at 
0.1~keV, 
Milky Way Halo (MWH) radiation with a temperature
fixed at 0.23~keV, 
and Cosmic X-ray Background (CXB).
%in the model. 
The LHB and MWH were modeled using
{\it apec} models
with a metal abundance fixed at the solar value (\cite{1989GeCoA..53..197A}),
and their normalizations were determined by fitting the
%the region excluding the point sources extracted from the 3XMM catalog and the CIZA1359 source. 
outermost region of the data after discarding sources 
that have fluxes of more than $1.0\times10^{-14}~\rm{erg~s^{-1}~cm^{-2}}$ in 0.4--7.2~keV using the ESAS {\it cheese} task, indicated by green circles in figure \ref{fig:fig1}.
\color{black}
%
%The CXB spectrum is a power law
The CXB spectrum is modeled with a power law
%,
of the photon index of $-1.41$.
Its normalization is
fixed at 
$5.7\times10^{-8}~\rm{erg~s^{-1}~cm^{-2}~st^{-1}}$ at 
2--10~keV (\cite{2009PASJ...61..339N}) 
because 
\color{black}
five Suzaku mapping observations showed that the ICM component filled the entire XMM field of view (\cite{KatoD}), and the CXB normalization cannot be estimated from the XMM data.
Strictly speeking, CXB normalization from \citet{2009PASJ...61..339N} is obtained by removing sources with flux exceeding $2.0\times10^{-14}~\rm{erg~s^{-1}~cm^{-2}}$, which differs from our case. However, as figure \ref{fig:fig1} shows, the ICM contamination make it difficult to remove all sources with the value. Also, even if we change the CXB normalization by 20\%, our results did not change within 0.3\% around the bridge, and 3\% at the outermost regions. Therefore, we continue using  $5.7\times10^{-8}~\rm{erg~s^{-1}~cm^{-2}~st^{-1}}$ at 2–10 keV as the CXB normalization, for simplicity.
\color{black}
%, and the photon index $\Gamma$ is fixed at $\Gamma=1.41$. 
A residual %SP
soft-proton (SP) 
contamination that is not filtered by the {\it pn filter}, was incorporated as a broken power law model with a cutoff of 3.0~keV (\cite{2008A&A...478..575K}).

%All these parameters are listed in Table 1.
%The source radiation model used the
To reproduce the source radiation, we used a
collisionally ionized diffuse
%gas 
thermal plasma 
model absorbed by the galactic absorption ($phabs$$\times$$apec$).
The redshift was fixed at 0.07 (\cite{2015PASJ...67...71K}). 
The metal abundance was fixed at 0.3 solar following \citet{2004PASJ...56..965F} and \citet{2021A&A...652C...3G} who used \citet{1989GeCoA..53..197A} abundance table. 
As a note, performing a spectral fit with \citet{2009LanB...4B..712L} abundance table as a trial does not significantly ($\sim$1\%) change the parameters of the source radiation.
\color{black}
The hydrogen column density was
fixed 
\color{black}
at $1.15\times10^{21}~\rm{cm^{-2}}$, determined from the HI map by the Leiden / Argentina / Bonn (LAB) survey (\cite{2005A&A...440..775K}).
The chi-squared statistic was used to evaluate the goodness of fit.
By fitting the spectra of each sub-region, we obtained a 
%足す
two-dimensional temperature (2D $kT$) map
(the upper left panel of figure \ref{fig:fig2}).
%% Fig にするなら、全てのところで Fig.XX にしてください。過去のPASJ論文参照のこと。

%%さすがに段落変えません？

%The 2D $kT$ map shows the existence of a hot region with temperature 7--8~keV in the bridge region indicated by the cyan dotted line in figure \ref{fig:fig2}.
%enhanced at high 
%The 2D $kT$ map shows the existence of a high temperature region in the bridge region.
%We defined this high-temperature region surrounding the boundary of the jump from 4--5 keV to 6--8 keV as indicated by the cyan dotted line in the 2 $kT$ map, as the ``hot region''.
The 2D $kT$ map shows the existence of a high temperature region in the bridge region.
We define this 6--8 keV high-temperature region, enclosed by the cyan dotted line in the map, as the ``hot region''.
It extends 
500~kpc 
\color{black}
along the merger axis. 
% We thought it made by two-shock waves heating. 
%There are temperature jumps at the southeast and northwest edges of the hot region.
%This suggests the presence of two shocks.
%
On the SE side boundary of the hot region, the temperature increases from $\sim$5~keV to $\sim$7~keV.
\color{black}
This jump is associated with a narrow enhanced emissivity
(see figure \ref{fig:fig1}).
Compared to the Suzaku results described by \citet{2015PASJ...67...71K}, the location of the temperature jump between their A2, A3 and A4 regions is consistent with that jump shown in our 2D $kT$ map obtained from the XMM EPIC-PN data.
\color{black}
The temperature values themselves are statistically consistent, as well (see figure 3 in \cite{2015PASJ...67...71K}). 
%The A3 region that it is high temperature compared to the other A2 and A4 given in figure 3 in \citet{2015PASJ...67...71K} is consistent with the location of this jump, as shown the 2D $kT$ map. 
%These results are roughly consistent with the hot-band region observed by Suzaku, as reported by \citet{2015PASJ...67...71K}.
%The Suzaku observations indicated the existence of a brightness discontinuity and a temperature jump (Kato et.al \cite{Kato+ 2015}). Its position coincides with the location of the temperature jump in the southeast of the 2D temperature map.
%
Furthermore, %a same 
a similar 
temperature jump 
%can be 
is observed at the NW
%side
edge
of the hot region, from $\sim$ 4 keV to $\sim$ 7 keV. 
%This is a new finding not found in Suzaku's observations. Both of these jumps are candidates for shocks.
%This was not resolved by Suzaku,
This was not well resolved in the first Suzaku observation (\cite{2015PASJ...67...71K}), probably due to its moderate angular resolution.
On the basis of their geometry, we considered these two edges as candidates for forward and backward shock waves.
\color{black}
%hot領域の南東側では5keV-->7keVに温度のジャンプが確認できる。すざくの観測から輝度不連続と温度ジャンプが確認されており、それは我々の南東の温度ジャンプの位置と一致している。
%加えて、hot領域の北西側にも同様の温度ジャンプが確認できる。これはすざくの観測では見つからなかった新たな発見である。この2つのジャンプは両方ともショックの候補である。

To better confirm 
these
shocks, we created the 2D pseudo electron density ($\bar{n}_{\rm e}$), pressure ($\bar{P}$), and ``astrophysical entropy'' ($\bar{S}$)
%map.
maps, also shown in figure \ref{fig:fig2}.
The $\bar{n}_{\rm e}$ was obtained from the normalization of the $\it{apec}$ model by assuming that the ICM is uniformly spread over 
%a
the 
line-of-sight depth of 1 Mpc and that the electron density $\bar{n}_{e}$ and hydrogen density $\bar{n}_{\rm H}$ satisfy the relation $\bar{n}_{\rm e}=1.2\bar{n}_{\rm H}$
considering the full ionization of helium.
The
$\bar{P}$ 
and 
%%% S
$\bar{S}$ 
were calculated as $\bar{P}=\bar{n}_{e}kT$ and $\bar{S}=\bar{n}_{e}^{-2/3}kT$, respectively. 
%Here, $\mu$ is the average molecular weight of the universe and was calculated to be 0.61. 
%From the puseudo density map (upper right panel of Fig.~2), we confirmed enhancement region in south east, so we decided temperature jump on southeast is a shockwave. We could not confirm enhancement region in northwest, but in the pseudo pressure map, the enhancement region in northwest is comfirmed and we decided it is newly finding shock wave on northwest.
%ここはまず圧力からかな、、、。
The 2D $\bar{P}$ map shows a clear enhancement at the SE edge. A similar enhancement is also marginally suggested at the NW edge. 
%From these enhancements in the 2D $kT$ and $\bar{P}$ maps, we concluded that there are two shocks at the edges of the hot region.
%(e.g. \cite{Simionescu2009}).
%密度は南東に見えるが北西には見えない。圧力では確認できるので北西の衝撃波と考える。
%chanel低密度を確認。密度からはマル。圧力では微妙(a little)だがエントロピーではエンハンスを確認-->channelと呼ぶ
%カラーバー縦。
%suggesting the existence of an equatorial shock.  However, the XMM-Newton point observation could not confirm the discontinuity due to insufficient field of view.

\begin{figure*}[ht]
 \begin{center}
 \includegraphics[width=17cm]{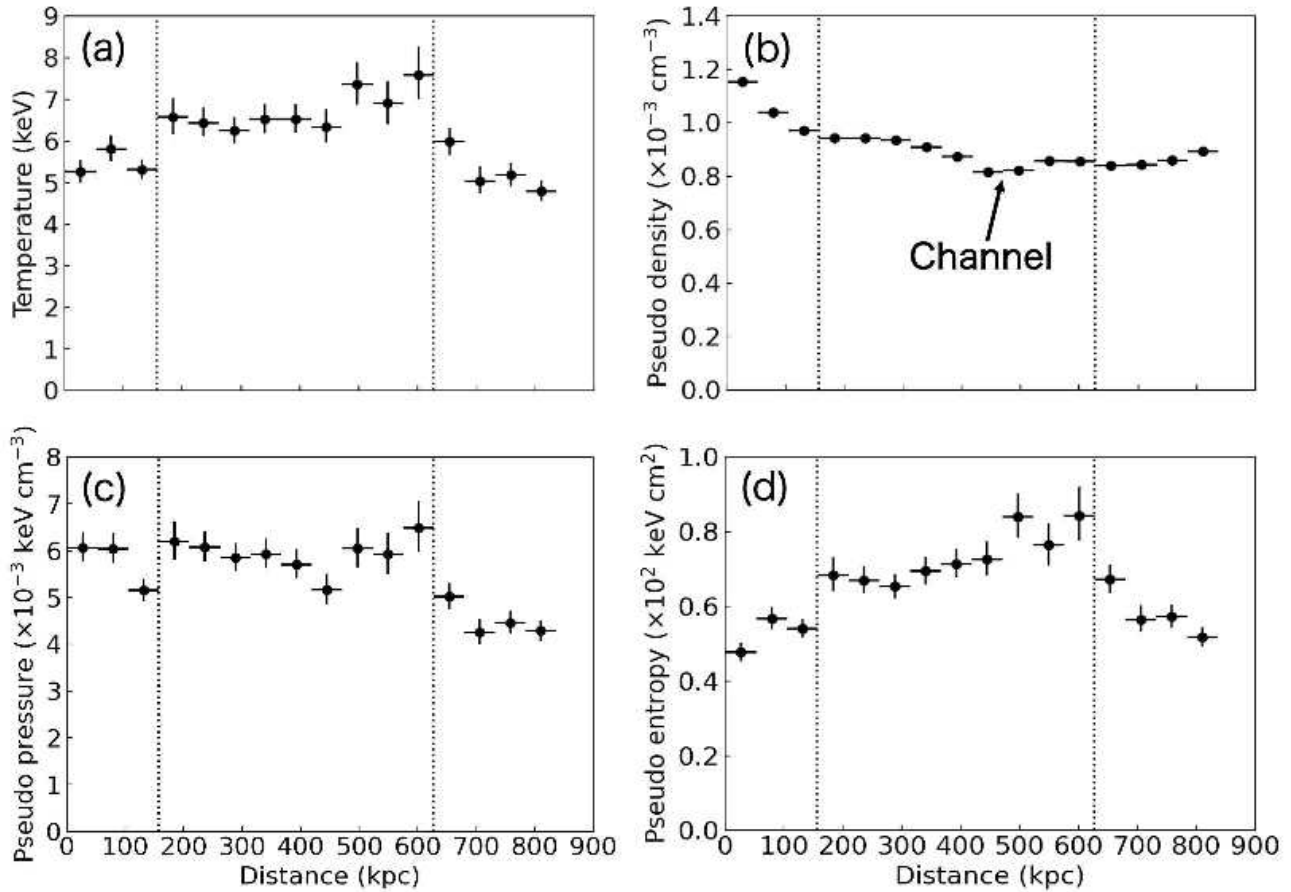}
 \end{center}
	%\begin{minipage}{0.33\hsize}
	%\begin{center} 
	%	\includegraphics[width=1.0\columnwidth]{figure/fig4_1_temperature_1d.eps}
	%\end{center}
	%\end{minipage}
	%\begin{minipage}{0.33\hsize}
	%begin{center} 
	%	\includegraphics[width=1.0\columnwidth]{figure/fig4_2_density_1d.eps}
	%\end{center}
	%\end{minipage}
	%\begin{minipage}{0.33\hsize}
	%\begin{center} 
%		\includegraphics[width=1.0\columnwidth]{figure/fig4_3_pressure_1d.eps}
%	\end{center}
%	\end{minipage}
\caption{1D temperature (a), pseudo density (b), pseudo pressure (c), and pseudo entropy (d) profiles from SE to NW. We divided 16 blue regions of 37".5 x 280" strips, as shown in the upper left panel of figure \ref{fig:fig2}. The vertical dotted--lines indicate the shock fronts at SE (left side) and NW (right side), respectively.}
%熱力学マップを図2の左上パネルのマゼンダラインで示す衝突軸に沿って南東から北西に投影した温度・pseudo密度・圧力1次元分布(左から右へ)。シアンの垂直線は南東(左)と北西(右)の衝撃波面。
\label{1d_profile}
\end{figure*}

\subsection{Estimating the ICM parameters around the two shock candidates}
\label{stimating the mach number of two-shocks}

%We projected a 2D temperature map to characterize the hot region. 
%To examine the jump condition, we created 1D profiles of these thermodynamic maps projected along the merger axis.
To investigate jump conditions, we created 1D thermodynamic profiles along the hot region.
For this purpose, the regionalization using the WVT algorithm is not appropriate because it is non-biased, based only on photon statistics, and not optimized to obtain ICM parameter differences perpendicular to the possible SE shock front observed in the X-ray image. We selected a rectangular region in the bridge as shown in the upper left panel of figure \ref{fig:fig2}. The region is then divided into 16 subregions, each with 36''.5 x 280'' area.  The 1D temperature profile thus obtained is shown in figure \ref{1d_profile}a. A pair of jumps in temperature are clearly visible. Similarly, $\bar{P}$ and $\bar{S}$ clearly show a pair of jumps in the same location (see figures \ref{1d_profile}c and \ref{1d_profile}d). The increase in pressure and entropy suggests that the two temperature jumps in the SE and NW are due to shock and not adiabatic compression. 
%We divided the 16 regions into strips from SE to NW, as shown by the magenta boxes in the upper-left panel of figure \ref{fig:fig2}, because statistical regionalization, such as the WVT binning algorithm, is not suitable to investigate specific structures.
%The 1D temperature profile shown in figure \ref{1d_profile}(a) indicates that there are clear temperature jumps on both shock fronts we defined.
%The 1D profile shown in figure \ref{1d_profile}(a) is a projection of the 2D {\it kT} map along the merger axis from the SE cluster center to the NW one (see the magenta line in the upper-left panel of figure \ref{fig:fig2}).
%Temperature jumps were indeed visible at the location identified as both shock waves, indicated by the dotted lines. 
\color{black}
%We projected the temperature map from the southeastern to the northwestern cluster center (see the magenta line in the upper left panel of figure \ref{fig:fig2}) as shown in 
%pink in Figure \ref{Thermodynamic_mapping} is shown in 
%figure \ref{1d_profile}(a). 

From these
% 足す
temperature jumps, the Mach numbers were estimated to be $M_{\rm{SE}}=1.25^{+0.15}_{-0.13}$ for the SE shock 
%wave
and $M_{\rm{NW}}=1.28^{+0.19}_{-0.17}$ for the NW shock \color{black}
%wave 
using the following Rankine--Hugoniot equation (R--H),
\begin{eqnarray}
\frac{kT_{\rm{2}}}{kT_{\rm{1}}}&=&\frac{5M^{4}+14M^{2}-3}{16M^{2}}.
%\nonumber
%\\
%\frac{n_{\rm{2}}}{n_{\rm{1}}}&=&\frac{4M^{2}}{M^{2}+3},
%\nonumber
\end{eqnarray}
Here, the subscripts 1 and 2 denote pre-shock and shocked conditions, respectively.
In this 1D analysis, we chose ``pre-shock '' and ``shocked'' regions as the ones closest to the shocks. 
The Mach number for the SE shock is consistent within error with $1.32\pm0.22$, the value obtained from the Suzaku observation
%The Mach number for the SE shock is within the error range of the Suzaku observation value $1.32\pm0.22$ 
%
(\cite{2015PASJ...67...71K}). 
\color{black}
%The projected distribution 2' (~150 kpc) away to the northeast is shown in blue line. The Mach number is estimated $M_{\rm{south}}=1.12^{+0.01}_{-0.01}$,$M_{\rm{north}}=1.14^{+0.03}_{-0.02}$.

On the contrary, no positive jumps in $\bar{n}_{e}$ are observed at the candidate shock wave locations, as shown in figure \ref{1d_profile}b. However, the ICM density of each cluster rapidly decreases with distance from the center, and a possible positive density jump could be smeared by its gradient. Therefore, the density distribution had better be compared to the pre-shock condition value. To address this possibility, we constructed 3D ICM profiles based on a few simple assumptions in the next section.
\color{black}
Focusing within the hot region, the $\bar{n}_{e}$ profile shows a two bins wide ($\sim$100 kpc) dip region in the center.
It is decreased by about 10\% compared to the neighboring regions, respectively. The feature could be caused by local shortage of depth of ICM emission, but more likely simply reflecting the density decrease.
Such a low surface brightness structure is called a channel, and a similar feature has been reported in some clusters, such as Abell 85 (\cite{2015MNRAS.448.2971I}).
Across the left boundary of the channel, the temperature has not changed within statistical error, but across the right side, it has increased by about 1 keV. Thus, the $\bar{P}$ profile shows a decrease in the left side and resumes in the right side. Similarly, the $\bar{S}$ profile shows almost no change in the left side and an increase by about 20\% in the right side. In short, the channel region has either lower ICM pressure or higher entropy. These features of channel would be caused by ``non-thermal pressure'' of magnetic or cosmic-ray origins or ``turbulence pressure''. It is also possible that it is affected by the high entropy ICM of the outer edge of the pre-collision cluster because this channel region is located roughly at the middle of the two clusters, and the contact surface (if not strongly mixed yet) of the two clusters will reside here. Although this is clearly an interesting signature, a lack of data statistics prevents us from further analysis.  In the 3D modeling presented later, to avoid the central regions, we will just focus on the thermal parameters at the edge of the hot region around the shock plane.

\color{black}
\color{black}

\section{Simple 3D modeling of the merger}

\subsection{The 3D original cluster model in the unshocked region}
%3D cluster modeling of pre-shock condition}
\label{3D modeling of pre-shock region}

The 2D $kT$ map and 1D profile strongly suggest that there are two shock waves in the bridge region.
\color{black}
This in turn means that the ICM in the ``outside regions'' is unaffected by the shocks, and record their pre-merger conditions. In addition, the two clusters are located about ~70\% of the $r_{200}$ of each other. Here, the $r_{200}$ estimated
from the temperature at the center of the cluster is 1.8 Mpc in
the NW and 2.1 Mpc in the SE cluster, respectively (\cite{KatoD}).
As such, ICM in the outside regions is not expected to be greatly affected by tidal forces. Here, for simplicity, we assumed that the two clusters are individually point-symmetric, and thus the ICM properties in the ``inside region'' before the collision can be estimated using a model parameterizing those in the outside regions. We modeled them three-dimensionally and hence call it a ``3D 
original cluster model''. 

\begin{figure}[tp]
 \begin{center}
 %%
 %%
 %\centering
  \includegraphics[width=8cm]{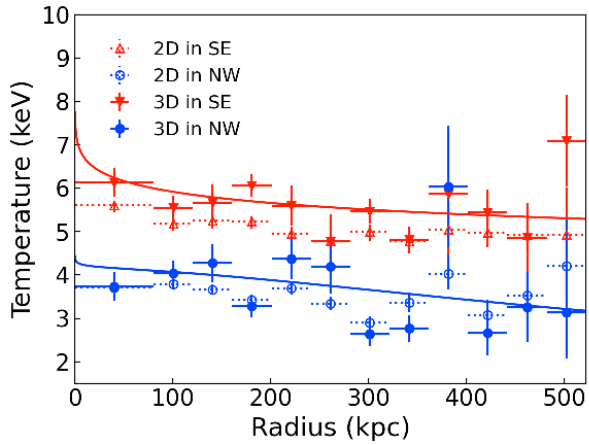}
 \end{center}
 \caption{The temperature distributions of the 3D original cluster model in the NW and SE clusters, respectively. Solid lines show the results of the fitting of the 3D temperature distributions in the NW and SE cluster with the 3D temperature model function given in \cite{2006ApJ...640..691V}, respectively.}
 \label{cold_temp}
%スペクトルを示す領域を付け加える
\end{figure}

To construct a temperature distribution of the 3D original cluster model, we selected  7'5 semicircular regions within the outside regions, indicated by green in the right panel of figure \ref{cold_temp}.
%The cold region used in density analysis slightly penetrates into the hot region 
%covers the southwest of the hot region 
%defined in the temperature map in the upper left panel of figure~\ref{fig:fig2}, and therefore, the semicircle regions in {\it SE} of $+138\degree\sim+318\degree$ and in {\it NW} of $-42\degree\sim+138\degree$ are 
%utilized here.
%selected as the cold temperature region.
%cold領域は図2の温度マップで定義したhot領域の南西と被っている、そのため半円領域の縮小した領域をcold温度領域として選択した
These semicircular regions were divided into annular sub-regions with radii of 0'--1',1'--1.5',1.5'--2',2.5'--3'$\cdots$,6.5'--7',7'--7.5' from each cluster center. 
The spectra of individual small regions were fitted to obtain the two-dimensional temperature radius distribution for the NW and SE clusters, respectively, as shown in figure~\ref{cold_temp}. 
\color{black}
%However, this temperature is a 2D temperature fitted to the spectrum along the line of sight. 
%\begin{figure*}[ht]
%	\begin{minipage}{0.5\hsize}
%	\begin{center} 
%		\includegraphics[width=1.0\columnwidth]{figure/fig5_1_3D_temperature.eps}
%	\end{center}
%	\end{minipage}
%	\begin{minipage}{0.5\hsize}
%	\begin{center} 
%		\includegraphics[width=1.0\columnwidth]{%figure/fig5_2_temp_rate.eps}
%	\end{center}
%	\end{minipage}
%\caption{Left: The temperature distributions of the 3D original cluster model in the NW and SE clusters, respectively. Solid lines show the results of the fitting of the 3D temperature distributions in the NW and SE cluster with the 3D temperature model function given in \cite{2006ApJ...640..691V}, respectively.  Right: A ratio map of the 2 $kT$ map and the emission-weighted temperature map based on the 3D original cluster model. The green semicircular region indicates the non-shocked region selected for the construction of the 3D original cluster model. The dotted cyan line indicates the hot region.}
%\label{cold_temp}
%\end{figure*}

Next, the spectra were de-projected using the xspec's {\it project} function to obtain the best-fit 3D temperature profiles.
%We
%
%then
%determined the best-fit 3D temperature 
%
%profiles
%by de-projecting the spectra using the xspec's {\it project} function.
%, as shown in magenta. 
%Resultant 3D temperature is shown in magenta plots.
%To do this, 
%Here, 
%we assume that the northern and southern clusters are spherically symmetric and that the ICM extends to 7'.5 ($\sim$600kpc) 
%due to 
%determined from 
%the observational field of view, respectively.
%In reality, as shown by the surface brightness in figure \ref{fig:fig1}, each cluster is 
%elliptical
%slightly elongated. 
%
%But 
%the temperature gradient is 
%not large, 
%modest so 
%
%we kept the circular assumption here, for simplicity.
%conclude that the assumption of spherical symmetry is not a problem. 
The outer two bins were excluded in the temperature distribution because they are considered 
%that the outer bin is overestimated and the temperature is dragged down.
to be deeply affected by the overly simplistic boundary condition of ignoring radii above 7'.5 (e.g., \cite{2005MNRAS.356..237J}; \cite{2008MNRAS.390.1207R}).
\color{black}
The 3D original cluster temperature model was then parameterized based on the following formula given by \citet{2006ApJ...640..691V},
%Some previous studies have used 
%the polytropic law~$T\propto \rho_{gas}^{\gamma-1}$ as a 3D temperature model. (\cite{M. Markevitch1999}; \cite{Finoguenov2001};\cite{G. W. Pratt2002}). However, the polytropic model has a problem of poor reproduction of the temperature profile at large radii. As a 3D cluster model for temperature, we adopted the expression for the outer part of the cooling core in \cite{Vikhlinin2005}, which solves this problem by increasing the number of degrees of freedom.
\begin{equation}
T(r)=t_{0}\frac{(r/r_{t})^{-a}}{[1+(r/r_{t})^{b}]^{c/b}}.
\end{equation}
%The large radius after 6.5' is represented by fitting a 3D galaxy cluster model to the 3D temperature distribution (magenta line in the figure\ref{cold_temp}). 
\begin{table}
\tbl{Parameters of 3D temperature original cluster model
%\footnotemark[$*$] 
}{%
\begin{tabular*}{8cm}{lccccc}
\hline
\begin{tabular}{l} \multicolumn{1}{c}{Region} \end{tabular} & \begin{tabular}{l} \multicolumn{1}{c}{$t_{0}$} \\ (keV) \end{tabular} & \begin{tabular}{l} \multicolumn{1}{c}{$r_{t}$} \\ (kpc) \end{tabular} & a & b & c \\
\hline
\multicolumn{1}{c}{NW} & $4.07\pm0.30$ & 85 &  0.01  &  0.3  &  0.1 \\
\multicolumn{1}{c}{SE} & $7.56\pm0.50$ & 425 &  0.01  &  2.0  &  0.5 \\
\hline
\end{tabular*}}\label{tab:first}
\begin{tabnote}
%\footnotemark[$*$] This list is just a sample. \\
%\footnotemark[$\dag$] Footnote for Value3.
\label{Parameters of 3D galaxy cluster temperature distribution}
\end{tabnote}
\end{table}
The best-fitting parameters 
for each
cluster are listed in table~\ref{Parameters of 3D galaxy cluster temperature distribution}.
%To evaluate the validity of the temperature distribution, we created %a 2D weighted by emissivity temperature map adapted WVT binning and divided this map by the 2D temperature map shown in the upper left panel of the Figure 2, as shown in the right panel of the Figure \ref{cold_temp}.

%This uncertainties are assumed to be the error in the constant term $t_{0}$ of the temperature model.

%As described in Section 2.2, the shock wave has not yet reached the cluster core, and the region which is not heated by the shock wave call the cold region. 
%To model the 3D ICM parameters in the pre-shock condition,
%We 
%we define the cold regions as a 
%circular sector region 
%circular sector regions in the {\it SE} cluster of $+108\degree\sim+348\degree$ and in {\it NW} of $-72\degree\sim+168\degree$ measured counter-clockwise from the west direction. %with the meger axis as the offset (Figure 1) and 
%
\begin{figure*}[ht]
	\begin{minipage}{0.5\hsize}
	\begin{center} 
		\includegraphics[width=1.0\columnwidth]{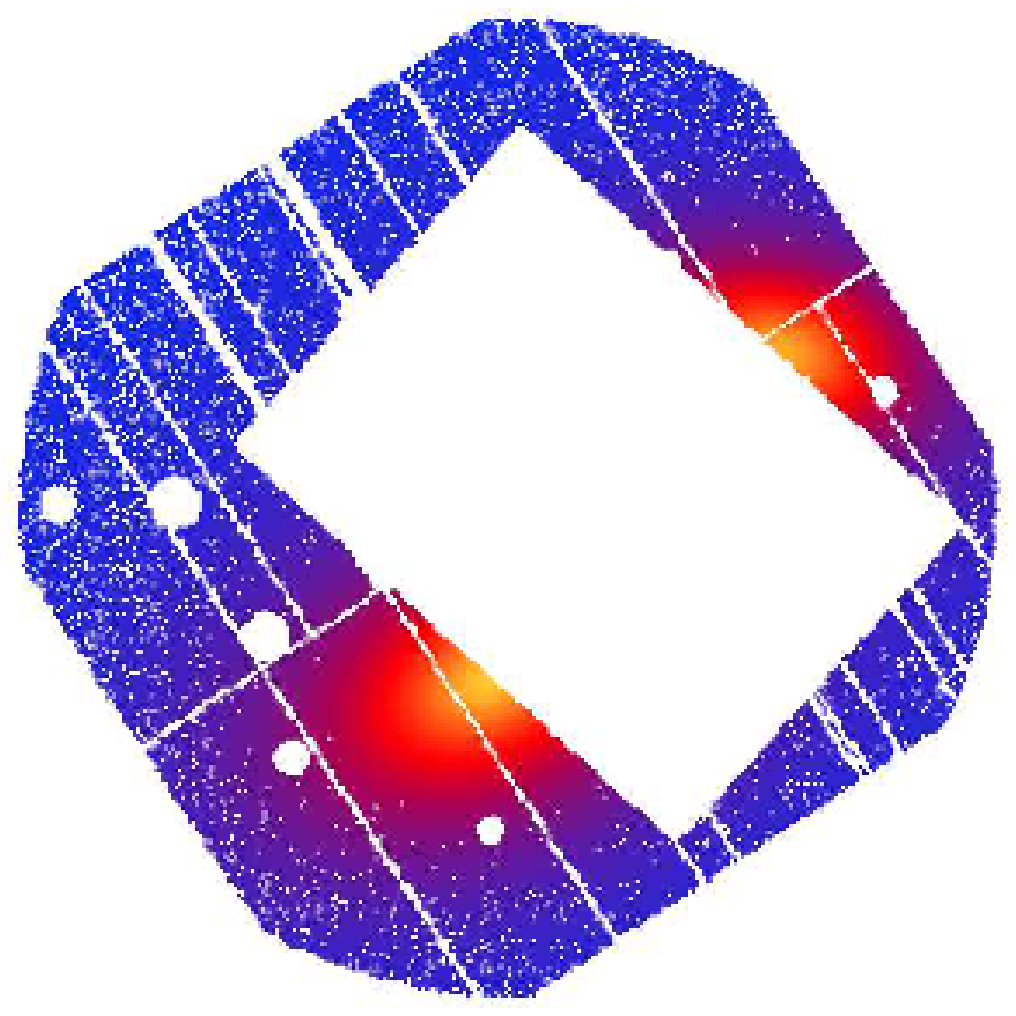}
	\end{center}
	\end{minipage}
	\begin{minipage}{0.5\hsize}
	\begin{center} 
		\includegraphics[width=1.0\columnwidth]{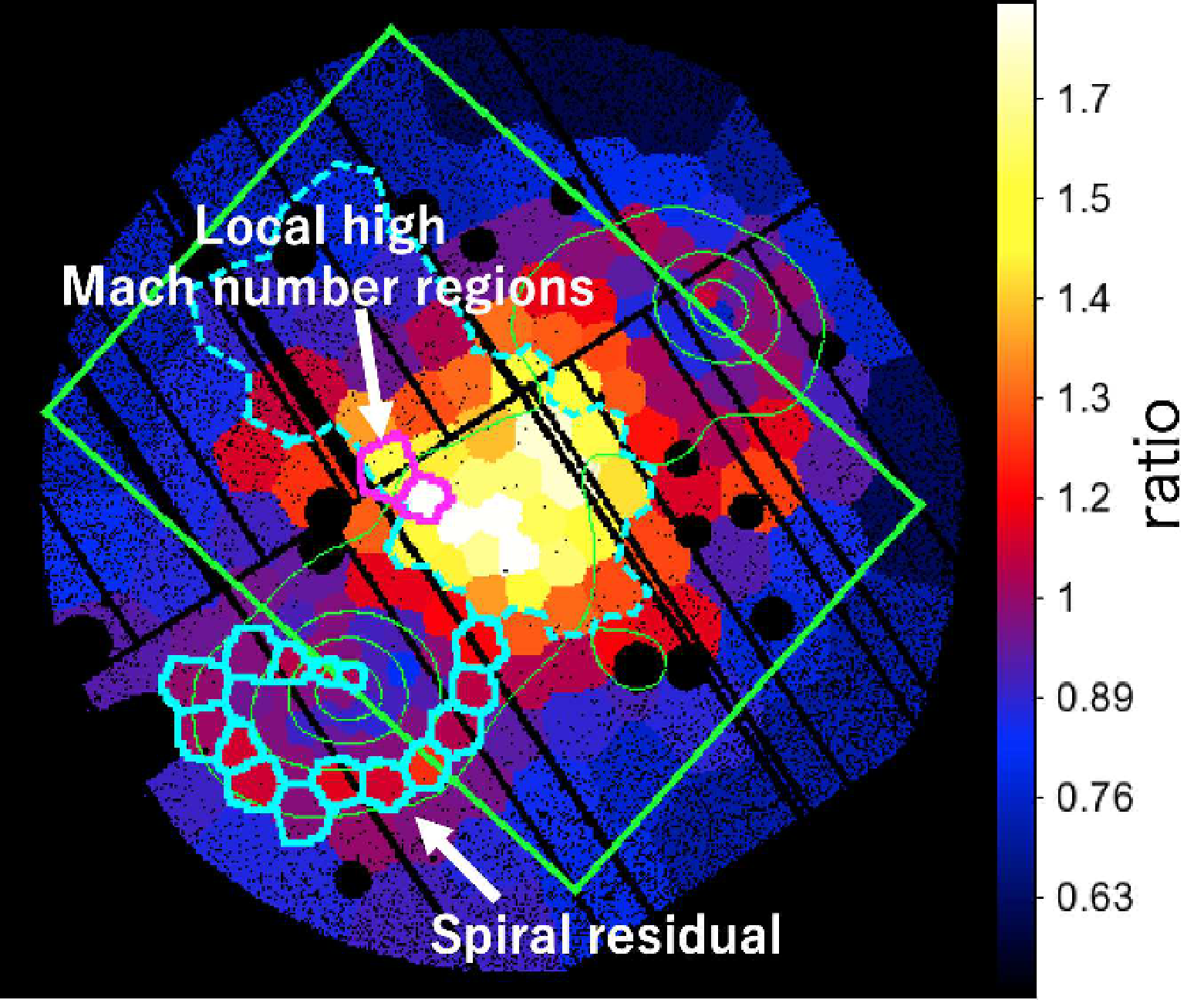}
	\end{center}
	\end{minipage}
\caption{Left: An image of the X-ray surface brightness model estimated from the two $\beta$ models. Areas excluded in the fitting process, such as the bridge region, are removed. The best-fitting parameters are listed in table \ref{tab:first}. Right: X-ray surface brightness ratio map of the model and observed images binned by the WVT algorithm. The green box indicates the areas excluded for the construction of the 3D original cluster density model. Cyan polygon regions are those with the spiral positive residual structure described in Section 3.5. The magenta polygon regions are the local regions with higher Mach numbers than the other regions indicated by the 3D merger model (see figure \ref{result}), corresponding to SE 6 and 7.}
\label{coldden}
\end{figure*}
Following $kT$ distribution,
\color{black}
a 3D 
%galaxy cluster 
ICM 
density distribution in the unshocked region was created.
%The gas distributions in the northern and southern clusters are assumed 
%to be according to 
%to follow 
%the elliptical $\beta$ model (\cite{Cavaliere1976}). 
%We created the 0.5--10~keV image removing 
%the $1.3--1.9~keV 
%to exclude the Al emission lines. 
We created 0.5--10~keV energy band image excluding the Al-line band (1.3--1.9~keV). 
The entire bridge region, indicated by a green box in the right panel of figure \ref{coldden}, point sources, and CCD dead spaces were also excluded.
%and excluded 1.3--1.9~keV energy band to exclude Al emission lines
%私たちはcold領域内の0.5~10keVイメージを作成しAl輝線を除くため1.3-1.9keVを除いた
The QPB component was subtracted from the created image, and an exposure correction was applied to that image.
%The QPB components were subtracted from the created image and an exposure correction was applied.
We executed simultaneous fitting on the created images with two $\beta$ model distributions of elliptical surface brightness and constant background component models for the NW and SE clusters ($\beta$~model$_{\rm{NW}}$ + $\beta$~model$_{\rm{SE}}$ + const).
%We executed simultaneously fitting to the created image with
%This image was simultaneously fitted with 
%the 
%two elliptical $\beta$ model
%distributions for the {\it NW} and {\it SE} cluster with a constant background component ($\beta$~model$_{NW}$ + $\beta$~model$_{SE}$ + const). 
%The 3D 
%galaxy cluster 
%ICM
%density distribution was obtained %to convert
%%
%by converting 
%the $\beta$~model derived from the surface brightness to a $\beta$~model of density, assuming the shape of the galaxy cluster to be an oblate ellipsoid.
For simplicity, the $\beta$ models obtained from the surface brightness fitting were transformed into $\beta$ density models assuming that the shape of the galaxy clusters was an oblate ellipsoid to obtain a 3D ICM density distribution.
\color{black}
The equation is
given by the following formulas:
\begin{eqnarray}
n(r)=n_{0}(1+(r/r_{c})^{2})^{-3\beta/2},\\
%r^{2}=\frac{x_{\rm{off}}^{2}(1-\epsilon)^{2}+y_{\rm{off}}^2}{r_{c}^{2}(1-\epsilon)^{2}},\\
r=[{x^\prime{}^{2}+(y^{\prime}}/\epsilon)^{2}]^{1/2},\\
x^{\prime}=(x-x_{0})~\rm{cos}~\theta~+(x-y_{0})~\rm{sin}~\theta,\\
y^{\prime}=(y-y_{0})~\rm{cos}~\theta~-(y-x_{0})~\rm{sin}~\theta,
\end{eqnarray}
where $x_{0}$ and $y_{0}$ are the coordinates of the center of the galaxy cluster, $\epsilon$ is the ellipticity (the ratio of the minor axis to the major axis), and $\theta$ is the angle of the major axis measured counterclockwise from the west direction.

\begin{table}[htbp]
\tbl{Parameters of 3D density original cluster model
%\footnotemark[$*$]
}{%
\scalebox{0.9}{
\begin{tabular}{lccccc}
\hline
\begin{tabular}{l} \multicolumn{1}{c}{Region} \end{tabular} & \begin{tabular}{l} \multicolumn{1}{c}{$n_{0}$} \\ $10^{-3}~\rm{cm^{-3}}$ \end{tabular} & \begin{tabular}{l} \multicolumn{1}{c}{$r_{c}$} \\ (kpc) \end{tabular} & $\beta$ & $\epsilon$ & \begin{tabular}{l} \multicolumn{1}{c}{$\theta$} \\ (degree)\end{tabular}\\
\hline
\multicolumn{1}{c}{NW} & $2.54\pm0.19$ & 165 & 0.67 & 0.68 & 348 \\
\multicolumn{1}{c}{SE} & $3.09\pm0.11$ & 188 & 0.53 & 0.66 & 26\\
\hline
\end{tabular}}}\label{tab:first}
\begin{tabnote}
%\footnotemark[$*$] This list is just a sample. \\
%\footnotemark[$\dag$] Footnote for Value3.
\label{Parameters_of_3D_density}
\end{tabnote}
\end{table}

The fitted model image is shown in the left panel of figure \ref{coldden} and individual best fitting parameters are shown in table~\ref{Parameters_of_3D_density}.
%We created a model image estimated from the $\beta$ models as shown the left panel of figure \ref{coldden}.
%The right panel of figure \ref{cold_temp} shows the ratio map, which divided this 2D emissivity-weighted temperature map adapted WVT binning by the 2D $kT$ map to evaluate the validity of the temperature distribution.
\color{black}
%a 2D emissivity-weighted temperature map, and adapted WVT binning and took the ratio to the 2D $kT$ map, as shown in the right panel of figure \ref{cold_temp}.
The right panel of figure \ref{coldden} shows a ratio map of observed and modeled images applied to the WVT binning algorithm to evaluate the residual of the surface brightness fitting.
%To validate the surface brightness fitting, WVT binning was applied to the corrected image 
%(after subtracting the the constant component)
%and the ICM model with two elliptical $\beta$ model images. 
%The right panel figure\ref{coldden} shows a map taking the ratio of the two images.
%The ratio of the two images was taken and presented in the right panel figure\ref{coldden}. %A ratio enhancement is present due to the emphasis on small variation, and it fall within the diamond-shaped area excluded by the fitting.
This ratio image shows that in the outer regions the model and observations mostly agree, while in the bridge region there is an increase in the ratio of more than 20\%.
The spread of the ratios in the outer regions was $\pm 15$\% for the NW and $\pm7.5$\% for the SE clusters, respectively. 
As the density fluctuation of the 3D original cluster model, the square root of these spreads was handled when evaluating the uncertainties of the parameters in the following analyses.

\begin{figure}[tp]
 \begin{center}
 %%
 %%
 %\centering
  \includegraphics[width=8cm]{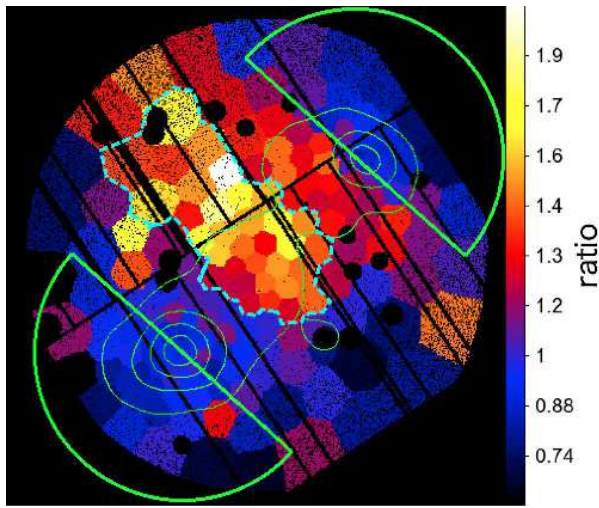}
 \end{center}
 \caption{A ratio map of the 2 $kT$ map and the emission-weighted temperature map based on the 3D original cluster model. The green semicircular region indicates the non-shocked region selected for the construction of the 3D original cluster model. The dotted cyan line indicates the hot region.}
 \label{temp_ratio}
\end{figure}
Similarly, we created the 2D emission-weighted temperature map by combining the 3D original cluster temperature and density models.
Our temperature model reproduces the observed 2D $kT$ map very well, except for the inside (or bridge) region.
Figure \ref{temp_ratio} shows a ratio map of the 2D $kT$ map and the 2D emission-weighted temperature map using the 3D original cluster model, with the WVT binning.
In the outside region, the deviations between model and observation were estimated to be 7.4\% for the NW and 6.4\% for the SE clusters, respectively. 
These uncertainties were included as the fluctuation of the 3D original cluster temperature model.

Overall, the enhanced region in the 2D $kT$ ratio map coincides well with the ``hot region'' on the 2D $kT$ map (the top left panel of figure \ref{fig:fig2}), as well as the enhanced region in the surface brightness ratio map. 
Actually, the spread of the surface brightness enhancement is more robust and smooth than the rather noisy enhancement in the 2D $kT$ map.
\color{black}
%In the un-shocked region, the error between surface brightness model and observation was estimated to be 15\% for the northwest and 7.5\% for the southeast, respectively.
%This uncertainties was included in the error in the constant term $n_{0}$ of the density model.
%These uncertainties are assumed to be the error in the constant term $n_{0}$ of the temperature model.

%A ratio map of surface brightness shows there is the clearly enhancement in hot region which can not be seen 1D pseudo density distribution (figure \ref{1d_profile}(b)).
%Here, we constructed the model consistent with the surface brightness enhancement by expanding the depth of the shocked condition, assuming that the un-shocked and shocked conditions overlap on the line of sight.
%Hereafter, we call it ``3D merger model''.
%A ratio is almost unity, expect for the regions not used for 3D pre-shock modeling (as shown as the blue diamond region).
%CIAO 4.13
%model (βmodel_north+βmodel_south+back(const)+NXB)*exposure

\color{black}

\subsection{The 3D merger model around the bridge region}
%3D merger modeling of the bridge region}
\label{3Dmodeling}

\begin{figure*}[htbp]
	%\begin{minipage}{0.5\hsize}
	\begin{center} 
		\includegraphics[width=16cm]{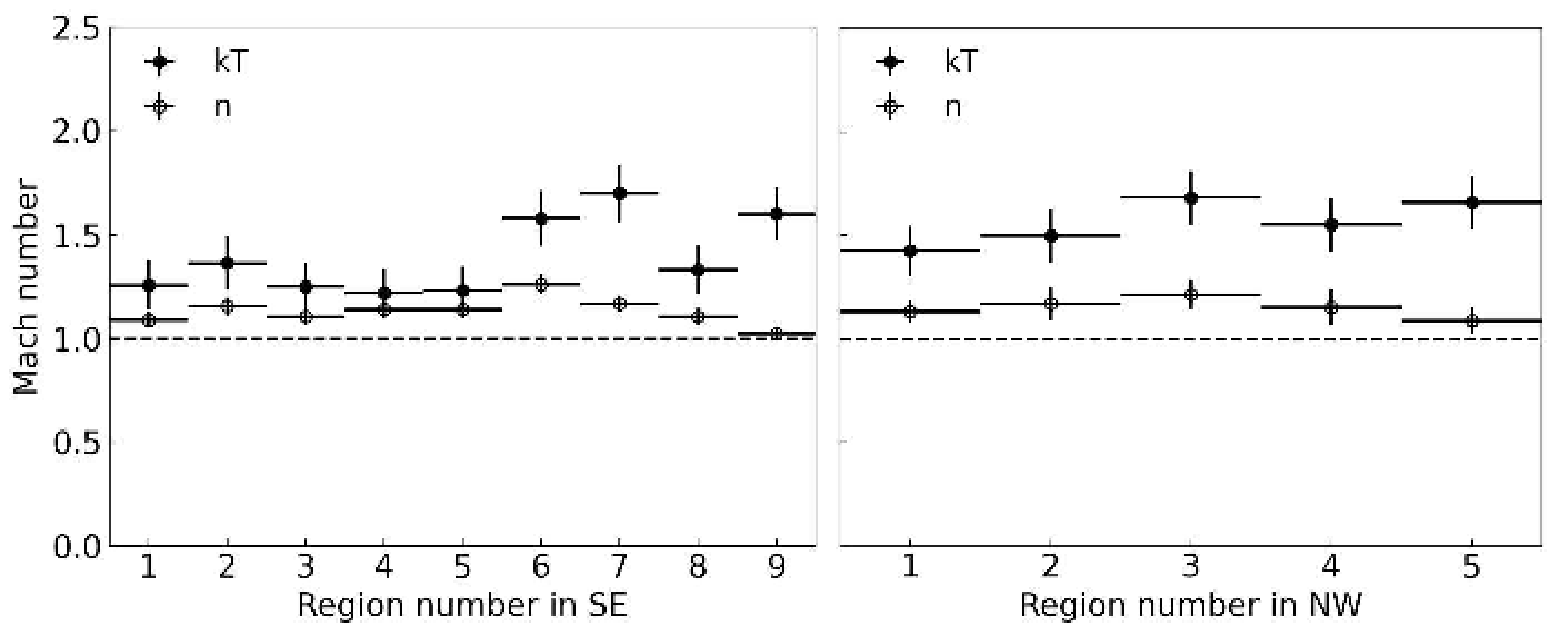}
	\end{center}
	%\end{minipage}
	%\begin{minipage}{0.5\hsize}
	%\begin{center} 
	%	\includegraphics[width=1.0\columnwidth]{1kT_mach_NW.pdf}
	%\end{center}
%	\end{minipage}
\caption{The Mach number distributions derived from the temperature (kT) and density (n) ratios at the SE (left) and NW (right) shock fronts. For simplicity, the density ratio is calculated as the square root of the surface brightness ratio.}
\label{fig:1kT_mach}
\end{figure*}

\begin{figure}[htbp]
 \begin{center}
 %\includegraphics[width=5cm]{pre_shock.png}
 %\end{center}
 %\begin{center}
 \includegraphics[width=7.5cm]{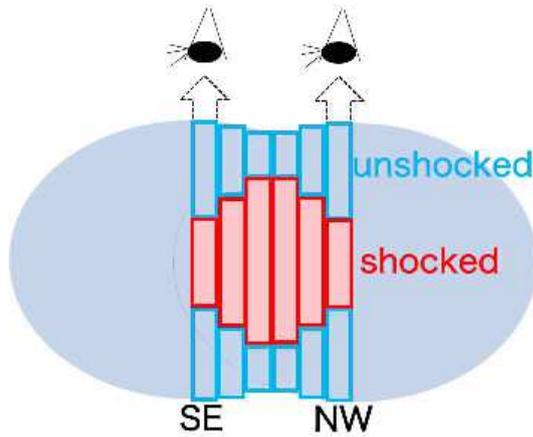}
 \end{center}
\caption{The concept of 3D merger modeling. The shocked and unshocked regions are indicated by red and  blue, respectively.}
%r200を追加・Hot領域全体ではなく衝撃波面が直方体の形をしていることを強調する
\label{model}
\end{figure}

%2.2章で短冊上に切って北と南の衝撃波のマッハ数を導出した結果温度・密度から出したマッハ数に違いが出た。密度は奥行き1~Mpcを仮定したが温度・密度のマッハ数が一致するときpre shock領域の奥行き~1Mpcに対し北と南の衝撃波面の奥行きは$l_{north}= \rm{Mpc}$,$l_{south}= \rm{Mpc}$となった。The result that the depth of the post shock was less than the depth of the pre shock did not indicate that the hot area was expanding in the line of sight, but rather that the two components, hot("shocked") and cold("unshocked"), were overlapping in the line of sight.Therefore,

\begin{figure*}[htbp]
 \begin{center}
 \includegraphics[width=15cm]{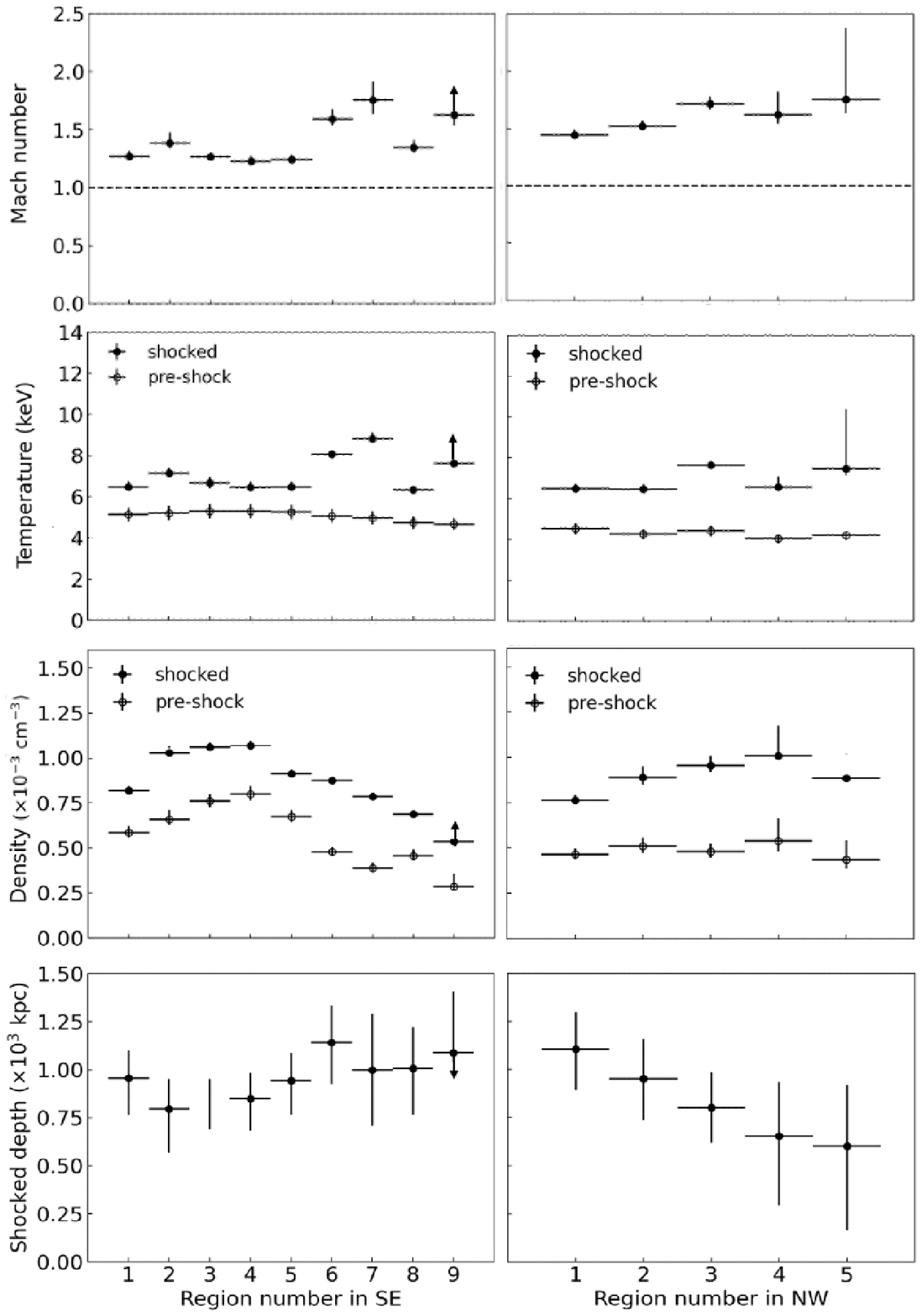}
 \end{center}
\caption{Results of the 3D merger modeling in the SE (left) and NW (right) shock fronts, respectively. From top to bottom, they represent the Mach number, temperature, density, and depth distributions.}
\label{result}

\end{figure*}

We successfully extracted the shock-wave enhancement region by restoring the 3D original cluster distribution. 
In other words, the apparent lack of density increase at the shock front in the 1D density profile (the upper right panel of figure \ref{1d_profile}) can be well explained as caused by the original ICM density gradient before merging.
\color{black}
Then, are the enhancement values of the temperature and X-ray surface brightness at the shock surface consistent?
We estimate the Mach numbers from the temperature and density ratio at the shock surface, respectively, using the R-H equation expressed in equation (1) and the following equation,
\begin{eqnarray}
\frac{n_{\rm{2}}}{n_{\rm{1}}}&=&\frac{4M^{2}}{M^{2}+3},
%\nonumber
\end{eqnarray}
where the subscripts 1 and 2 indicate pre-shock and shocked conditions, respectively.
For simplicity, the density ratio is calculated as the square root of the X-ray surface brightness ratio.
We defined the shock front candidate regions as SE 1–9 and NW 1–5, located at the edge of the hot region, as shown in the upper right panel of figure \ref{fig:fig2}.
Figure \ref{fig:1kT_mach} shows the Mach numbers estimated from the temperature and density ratios for individual regions.
Overall, the Mach numbers for the temperature are clearly larger than the Mach numbers for the density.
%In some regions of the SE shock front (e.g., Region SE 3), the Mach numbers of the two coincide within a range of error, but they differ by $\sim$ 8\% comparing with best-fit values .
%In some regions of the SE shock front (e.g., Region SE 3), the Mach numbers of the two coincide within a range of error, but 
%An F-test also indicates that the two Mach numbers do not agree in the 3$\sigma$ confidence level ($F=9.5>F_{\rm{3\sigma}}=2.6$).
%We projected the temperature and density ratio map as in section 2.3 from the SE cluster center to NW one center, as shown in figure \ref{fig:ratio_profile}, respectively.
%For simplicity, the density ratio is taken as the square root of the X-ray surface brightness ratio.
%Outside of the shock wave surface, the value is almost 1, i.e., the model and observed values are in agreement. On the inner side, a clear enhancement can be confirmed.
%Outside of the shock  surface, the model and observed values are in agreement and, on the inner side, a clear enhancement can be confirmed. 
%According to equation 1, the Mach numbers on the SE and NW shock fronts are calculated to be $M_{\rm{SE}}=1.22~\pm~0.11$ and $M_{\rm{NW}}=1.50~\pm~0.12$, respectively.
%On the other hand, from the density ratio, the Mach numbers are calculated to be $M_{\rm{SE}}=1.14~\pm~0.04$ and $M_{\rm{NW}}=1.17~\pm~0.07$ according to the following equation,
Because the Mach numbers obtained from temperature and density jumps are inconsistent, it is likely that the shocked and unshocked components overlap on the line of sight within the hot region.
\color{black}
Therefore, we tried to fit the spectra with a two-temperature (2 $kT$) model to derive the thicknesses in the line of sight of the shocked and unshocked components, $l_{\rm s}$ and $l_{\rm us}$, respectively. However, as discussed in the Appendix, neither the shocked temperature ($kT_{\rm s}$) nor the $l_{\rm s}$ can be determined simply by fitting the 2 $kT$ model with both parameters set free.

Here we introduced another simple
assumption that the shocked and unshocked components have the 3D-special structure shown in figure \ref{model}.
\color{black}
The ICM parameters for the unshocked component can be estimated from the 3D original cluster model. 
The preshock condition, which is the state in the shocked region if the shock wave did not pass through, can also be estimated using the same model, and the R-H relations shown in equations (1) and (7) are assumed between the shocked and preshock components on the shock wave surface.
Precisely speaking, this assumption is valid only at the edge region of the shock because the post-shock region is already compressed, and thus its pre-shock condition is difficult to estimate. We therefore focus on the shock front candidate regions for a moment.
\color{black}
As a reminder, we note here that the subscripts ``s'', ``un'', and ``pre'' labels mean those for the shocked condition, unshocked condition, and preshock condition of the shocked region, respectively.
%The subscripts ``s'' labels imply those for the shocked region, ``un'' for an un-shocked condition, and ``pre'' for a pre-shocked condition of the shocked region, respectively. 
The derivation of the $l_{\rm s}$ and Mach numbers proceeded specifically as follows.

\begin{enumerate}
\item Assuming $l_{\rm s}$, the $norm_{\rm us}$ and $kT_{\rm us}$ of the unshocked component can be calculated from $l_{\rm us}$ and the 3D original cluster model by emission weighting (here ICM was considered up to $r_{200}$).

\item With the low temperature component, $kT_{\rm us}$ and $norm_{\rm us}$, fixed at this calculated value, a spectral fit of the two temperature model was performed to derive $norm_{\rm s}$ and $kT_{\rm s}$ for the high temperature component.

\item Similarly, $n_{\rm pre}$ and $kT_{\rm pre}$ for the preshock condition can be calculated from the 3D original cluster model by weighting. Then, by applying the R-H relation for the density and the temperature, two estimations of Mach number can be calculated.

\item After repeating this calculation for various $l_{\rm s}$, the depth $l_{\rm s}$ was derived by finding the point where the two Mach numbers are consistent.

\end{enumerate}
Details are given in the Appendix.
The results are %shown
summarized in figure \ref{result}. The Mach numbers are about $\sim$1.3 for the SE shock surface (excluding the region number 6--7) and $\sim$1.7 for the NW shock surface. In both cases, $kT_{s}$ is $\sim$ 8 keV, $n_{s}$ is $\sim 1.0\times10^{-3}~\rm{cm^{-3}}$ and $l_{s}$ is $\sim$ 1~Mpc. 
\color{black}

\subsection{Comparison with Planck SZ signal}

%From 
%
%\S 
%subsection
%\ref{3Dmodeling}, 
%we found that the shock planes have a 
%hot component of 
%a post-shock region sandwiched with two shock with Mach number $\sim$ 1.7 in {\it NW} and $\sim$ 1.3 in {\it SE}, with a depth of $\sim$ 1~Mpc.
%Here we quantify the entire post-shock condition, assuming that the {\it SE} cluster has a Mach 1.3 reverse shock wave and the {\it NW} cluster has a Mach 1.7 reverse shock wave, and that the two shock waves are generated at the center of the hot region.
%セクション3.3から我々は3次元モデルを衝撃波面に適応することによって北西と南東の衝撃波はそれぞれマッハ~1.7と~1.3でどちらとも~1Mpcの奥行きを持つことを発見しました。ここで我々はpost-shock領域全体の状態を定量化するため、この2つの衝撃波はpost-shock領域の真ん中で発生し、1Mpcの奥行きを持った各銀河団の逆行衝撃波がICMに影響を及ぼしているモデルを想定しました。
%ここで我々は南東の銀河団はマッハ1.3の逆行衝撃波を北西の銀河団はマッハ1.7の逆行衝撃波を持ち、どちらともhot領域の中心で発生し、hot領域全体を表現した
%ここで我々は衝撃波はHot領域の中心で衝撃波が発生し、南東と北西の衝撃波はマッハ1.3と1.7の逆衝撃波としてそれぞれの銀河団に影響を及ぼしているとし、hot領域全体のモデルを仮定した
%
%We verify whether this assumed model is correct or not using the Sunyaev-Zel'dovich (SZ) effect signal.
In this section, we verified the 3D merger model with the Sunyaev-Zel'dovich (SZ) effect signal using the Planck satellite observations.
\color{black}
%In this subsection, we compared the Sunyaev-Zel'dovich (SZ) effect signal predicted from our 3D merger model and that observed by the Plank satellite.
%we found that the shock planes have a 
The 3D merger model showed that the shock planes have
a post-shock region sandwiched between two shocks with Mach numbers $\sim$ 1.7 in NW and $\sim$ 1.3 in SE, with a depth of $\sim$ 1~Mpc.

\begin{figure*}[t]
	%\begin{minipage}{0.5\hsize}
	\begin{center} 
		\includegraphics[width=16cm]{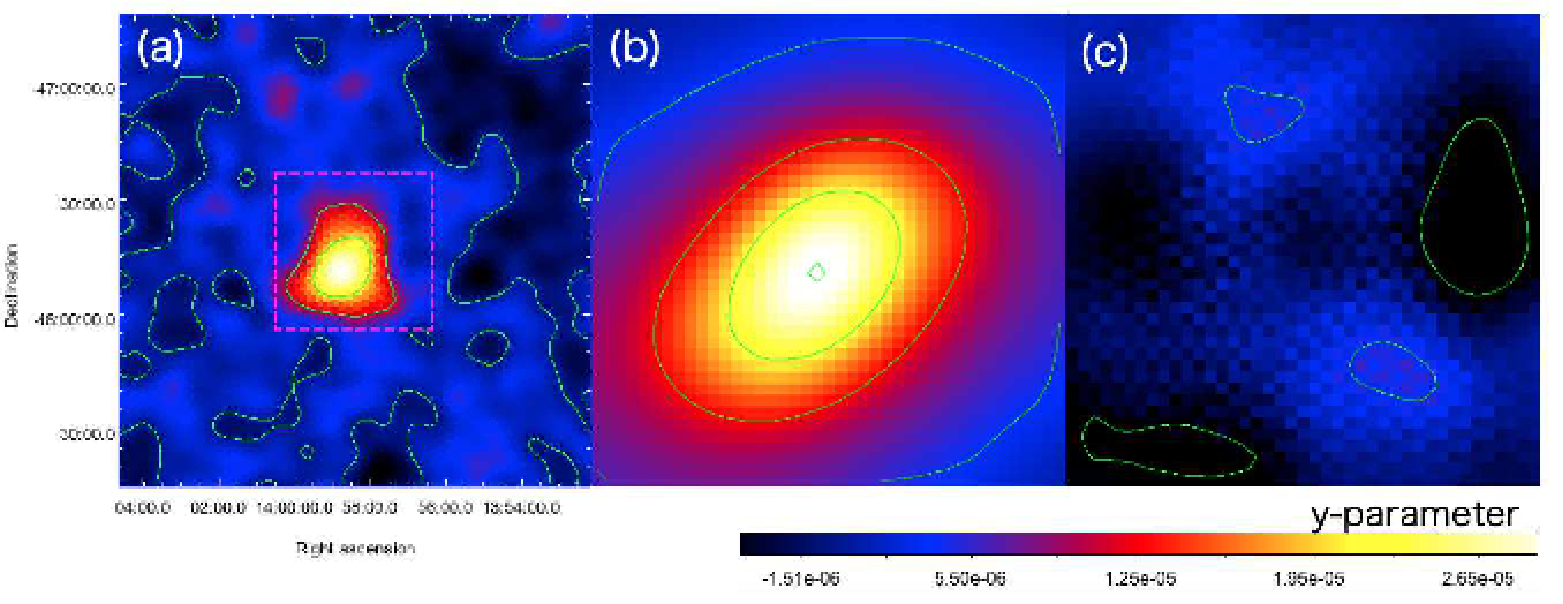}
    \end{center}
	%\end{minipage}
		%\begin{minipage}{0.33\hsize}
	%\begin{center} 
		%\includegraphics[width=1\columnwidth]{Planck_ymap_overall.pdf}
	%\end{center}
	%\end{minipage}
	%\begin{minipage}{0.33\hsize}
	%\begin{center} 
		%\includegraphics[width=1\columnwidth]{model_ymap.png}
	%\end{center}
	%\end{minipage}
	%\begin{minipage}{0.33\hsize}
	%\begin{center} 
		%\includegraphics[width=1\columnwidth]{Planck_model_ymap_residual.pdf}
	%\end{center}
	%\end{minipage}
\caption{Left: The ymap of the square region $123' \times 123'$ centered on CIZA J1359 provided by the Planck Release2 NILC ymap (\cite{2016A&A...594A..22P}). The magenta dotted line indicated the $41' \times 41'$ square region. Center: The ymap based on the X-ray 3D model within the magenta-dotted region. It has an angular resolution of 10'. Right: Residual map of the center figure subtracting the left figure. The color bar and green contour indicate the y-parameter value, which is dimensionless in units, and the intensity level of $3 \times 10^{-6}$.}
\label{fig:sz}
\end{figure*}
The combination of X-ray surface brightness and the Compton y-parameter gives us information 
about depth and electron density because of their different dependence on electron temperature and density (e.g., \cite{2017A&A...606A...1A}; \cite{2022MNRAS.510.3335H}). 
\color{black}
%The Compton y-parameter is given by the following %tabular expression:
%equation; 
%\begin{eqnarray}
%y=\int{\frac{k_{b}T}{m_{e}c^{2}}n_{e}\sigma_{T}dl}\nonumber
%\end{eqnarray}
%where $k_{b}$ is Boltzmann's constant and $\sigma_{T}$ is the Thomson scattering cross section. 
%We verified the validity of the 3D model by checking whether the y calculated from the model agrees with the y actually observed. 
We use the ymap obtained using the Planck Release2 Needlet Independent Linear Combination (NILC; \cite{2009A&A...493..835D}; \cite{2013MNRAS.430..370R}) method.
%We used the observation data from the Planck Release2 Needlet Independent Linear Combination (NILC; \cite{2009A&A...493..835D}; \cite{2013MNRAS.430..370R}) ymap provided by the Planck collaboration (\cite{2016A&A...594A..22P}).
Figure \ref{fig:sz}a shows the ymap provided by the Planck collaboration (\cite{2016A&A...594A..22P}) of the 123' $\times$ 123' square region centered on CIZA J1359, $(\alpha, \delta)_{\rm J2000.0} = ($\timeform{13h58m30s.0}, $-$\timeform{47D45'00''.0}).
We note here that the ymap produced by NILC is convolved with a Gaussian kernel with FWHM=10', and therefore it is impossible to directly compare it with our X-ray maps.
In the 41' $\times$ 41' square region of figure \ref{fig:sz}a indicated by the magenta dotted line, we calculated the y-parameter based on the 3D merger model and convolved it with the same FWHM as shown in figure \ref{fig:sz}b.
For simplicity, the post-shock ICM condition is assumed to be constant from the shock front to the contact surface (at the middle of the channel. See section 2.3) and spread over a depth of 1 Mpc.
\color{black}
Although being too much simplified, the density and temperature distribution shown in figure \ref{1d_profile} is largely consistent with this assumption.
To check the consistency of the model with the observed data, we created the residual map by subtracting the calculated map from the observed map as shown in figure  \ref{fig:sz}c.
The residual difference is about $3.0\times10^{-6}$, which is consistent with the CMB fluctuations.

These results mean that the 3D merger model with simple assumptions for the shocked condition is consistent with the y-paramter observed by Planck within the CMB fluctuations.
However, our analyses here are widely smoothed and the resultant image cannot even resolve the two cluster cores.
%In principle, comparing X-ray and SZ signals will resolve $l_{s}$ if we assume a relatively simple geometry (e.g. \cite{2017A&A...606A...1A}; \cite{2022MNRAS.510.3335H}). However,
The current angular resolution of the SZ data does not allow for further investigation of this specific merging cluster.

\color{black}

\section{Discussion}

\subsection{The shock velocity and age inferred from the 3D merger model}

In the SE shock, the preshock temperature $kT_{\rm{pre}}$ = $5.3^{+0.3}_{-0.3}$ and the Mach number of $\sim$ 1.3 derived from our model provide the shock front (preshock) velocity of $\sim 1600~\rm{km~s^{-1}}$, and the postshock velocity of $\sim 1000~\rm{km~s^{-1}}$. 
We converted these velocities from the shock front frame to the postshock frame, i.e., subtracting the postshock velocity from the velocities of all other regions. 
Based on the postshock frame, the shock front is moving with a speed of $\sim 1000~\rm{km~s^{-1}}$, while the preshock ICM (of the SE cluster) is approaching at $\sim 600~\rm{km~s^{-1}}$ (see figure 9.2 in \cite{KatoD} for a schematic explanation about this calculation process). 
\color{black}
Similarly, in the NW shock, the shock front is proceeding at a speed of $\sim 900~\rm{km~s^{-1}}$, while the preshock ICM (of the NW cluster) is approaching at $\sim 900~\rm{km~s^{-1}}$. 
Combined, the post-shock region is estimated to be expanding at the velocity of $\sim 1900~\rm{km~s^{-1}}$ and two clusters are estimated to be merging at the velocity of $\sim 1500~\rm{km~s^{-1}}$ on the sky plane.
The width of the hot region on the merger axis shown in the upper left panel of figure \ref{fig:fig2} is 
$\sim 500$~kpc.
\color{black}
If the shock waves pass through this width at a constant speed, the age of the shock wave is estimated to be 
$\sim 260~\rm{Myr}$.
\color{black}
%個々1シグマで誤差出して合計誤差を求める

%\begin{figure}[h]
 %\begin{center}
 %\includegraphics[width=8cm]{shock_velocity.png}
 %\end{center}
%\label{shock_velocity}
%\caption{The illustration of velocity in the pre-shock region (blue), post-shock region (red), and shock wave (gray) in the shock wave system (left) and post-shock system (right).}

%\end{figure}

%\subsection{Expectations for other observations}
Based on these velocities and Mach numbers,
the kinetic energy flow through the SE shock front can be estimated using the 
following equation (e.g., \cite{2019NatAs...3..838G}; \cite{2019MNRAS.488.5259Z}),
\begin{eqnarray}
F=\frac{1}{2}\rho_{\rm{pre}} v^{3}(1-\frac{1}{C^{2}})S,\nonumber\
\end{eqnarray}
where $\rho_{\rm{pre}}$ is the preshock ICM mass density \footnote{We calculated $\rho_{\rm{pre}}$ as $7/6m_{\rm H}n_{\rm pre}$, assuming a hydrogen to helium mass ratio of 1:4 and a density ratio of 10:1.}, 
$C$
is a compression number calculated as the ratio of the preshock ICM mass density to the shocked one, 
and
$S$ is the surface area of the shock. 
%$\rho_{\rm{pre}}$ is calculated as $7/6m_{\rm H}n_{\rm pre}$ based on the assumption that the relation $n_{\rm e}=1.2n_{\rm H}$ is satisfied.
From 
the 
3D merger model, 
the SE shock front has a surface area of $1~\rm{Mpc}\times1~\rm{Mpc}$ with a Mach number of 1.3 ($C\sim1.4$) and a velocity of $1600~\rm{km~s^{-1}}$, and $n_{\rm pre}$ is $6.0\times10^{-4}~{\rm{cm^{-3}}}$. With these parameters, the kinetic energy flow through the shock front is calculated to be $\sim~2.2\times10^{45}~\rm{erg~s^{-1}}$. 

%Figure 16 in Paul's paper, which simulates a cosmological merger, shows that about 10% of the total pressure exerted by the shock wave is dissipated in the turbulence of the ICM.
\citet{2011ApJ...726...17P} simulated the level of turbulence in a cluster merger and suggested that about 10\% of the shock kinetic energy 
would 
\color{black}
be converted into turbulence in the post-shock region. If this number is applied to the SE shock of CIZA J1359,
%Figure 16 in \citet{2011ApJ...726...17P}, which simulates a cosmological merger, shows that about 10\% of the total pressure imparted by the shock wave is dissipated into the turbulence of the ICM. Assuming that 10\% of the total kinetic energy of the southeastern shock wave is utilized for ICM turbulence,
the line--of--sight turbulent velocity is calculated to be $\sim 200~\rm{km~s^{-1}}$. %198 km s~. 
\color{black}
This value indicates that XRISM {\it{Resolve}} observations of CIZA J1359 could potentially be able to distinguish turbulent velocities in the shocked and unshocked regions.

More recently, \citet{Kurahara2022} discovered 
a candidate for 
\color{black}
diffuse cluster radio emission in the bridge region using uGMRT and MeerKAT radio observatory data. It is located closely at the NW shock front with a Mach number of 1.7.
\color{black}
Meanwhile, they did not find any other diffuse cluster radio emission symptom.
If there is a diffuse radio source associated with the higher Mach number shock, it can be relic high-energy electrons reaccelerated by Fermi first-order acceleration. The lack of diffuse radio sources in other post-shock regions suggests that the Fermi second-order electron acceleration is in-efficient in these young shock regions. A comparison between the data of future XRISM and SKA (Square Kilometre Array) would reveal the nature of turbulence and turbulent acceleration in detail.
\color{black}
%preshock乱流0km/s
%Assuming in reg~28 turblence velocity $198~km s^{-1}$ and 3D modeling parameter, we made simulated spectrum observed by XRISM $resolve$ with exposure 300~ks using xspec $fikeit$ comamand. The response files and background file got from XRISM researcher page. We fited this spectrum using vapec model with the tublence velicity and redshift free ten times and the turblence velocity is $196^{+46}_{-48}~\rm{km~s^{-1}}$
%cold領域と区別できるか
%1.ペルセウス銀河団(160km/s)をover estimateしたcold領域と仮定して話をする。
%2.シミュレーションから銀河団の典型的な乱流速度を求めた論文を引用する。
%Nagai et.al 2007のreferしてシミュレーションとモデルを比較

%誤差3つ確かめる。pre shock温度
%衝撃波速度vを用いて衝撃波の与えるエネルギーフラックスは式で与えられる。$\rho$は数密度、
%XRISMより乱流のエネルギーが計算できる。もし運動エネルギーの10\%が乱流に使われたとすると(ここで参考文献を引っ張ってきたい)観測されるFe輝線幅は$\cdots$になる。
%Nicerで観測した時に高エネルギー側で2温度がどのように見えるか。さらには分解できるか
%衝撃波の運動エネルギーの10\%が乱流に使われたとすると乱流速度は381~km/sと計算できる。また、この乱流を高エネルギー分解能の検出器で観測した場合、FeのHe-$\alpha$はFWHM=8.37~eVの輝線幅を持つ。また、検出器のエネルギー分解能や鉄イオンが熱運動する幅を考慮するとFWHM=10.0~eVとなった。検出器のエネルギー分解能はひとみのSXSと同じFWHM=4.9~eVとした。

%S.Paul+ 2018のシミュレーションでは衝撃波が生成されてから2Gyr以内では衝撃波が与えるトータルプレッシャーの内10\%以上が乱流に使用されている。

\subsection{Possibility of local velocity structure in the SE cluster}

On the SE shock, \citet{KatoD} reports a linear brightness enhancement using the Chandra data, which is apparent to the north-east and gradually fades out to the south-west. A similar trend is also visible in our data (see figure \ref{fig:fig1}), and SE 6 and 7 are among the brightest regions in the enhancement. This local linear enhancement was interpreted as a 70 kpc wide ``narrow shock'', which has just emerged, in \citet{2015PASJ...67...71K} and \citet{KatoD}. 

The 70 kpc narrow shock associated with the southern end of the
500 kpc 
\color{black}
post-shock region can be consistently understood if there are local fluctuations in the ICM condition, such as local bulk motion. 
\color{black}
Using the preshock temperature of the SE shock and the derived Mach number at SE 6 and 7, the preshock velocity observed from the postshock region is estimated to be $\sim 1100~\rm{km~s^{-1}}$. This is $\sim 500~\rm{km~s^{-1}}$ faster than those estimated in other SE regions. In other words, if there is a local bulk velocity of $\sim 500~\rm{km~s^{-1}}$ in this region, the enhancement in SE 6 and 7 regions (and also in nearby regions) can be understood. 

In the surface brightness ratio plot (the right panel of figure \ref{coldden}), there is a positive residual of 10-20\% levels with spiral shape. The residual is also pointed out by \citet{KatoD} as a surface brightness jump in the Chandra image, which is interpreted as a cold front candidate. If there was a small galaxy group infalling into the southern cluster from the south-east and then curving to the south-west prior to the major merger, residual and local velocity enhancement could take place. Although a more detailed discussion is not possible with the current limited information, a $\sim 500~\rm{km~s^{-1}}$ local velocity structure is suggested for a similar spiral residuals by, e.g., \citet{2019ApJ...871..207U}, and in numerical simulations (e.g., \cite{2006ApJ...650..102A}). \citet{2020A&A...633A..42S} also detected signatures of local velocity enhancement of an order of $\sim 500~\rm{km~s^{-1}}$ to $\sim 1000~\rm{km~s^{-1}}$ in the Perseus and Coma clusters of galaxies. As such, the $\sim 500~\rm{km~s^{-1}}$ local velocity structure around SE 6 and 7 is possible.
\color{black}
%This velocity is %in good agreement with that 
%within the range of those 
%expected by gas sloshing in numerical simulation and observation (e.g. \cite{2006ApJ...650..102A} \& \cite{2019ApJ...871..207U}).

\section{Conclution}

With a 65~ks XMM--{\it Newton} EPIC--PN data, we studied the nearby, merging galaxy cluster CIZA J1358.9$-4750$.
The main results of our work are summarized in the following.
\begin{enumerate}[(i)]
 \item The 2D thermodynamic maps, segmented into $\sim$ 180 regions, indicated the existence of a 7--8~keV high temperature region that extends 500 kpc \color{black} along the merger axis. The clear increase in temperature, pseudo-pressure and pseudo-entropy \color{black} at the edges of the hot region led us to conclude that there were two shock waves, one in the SE and the other in the NW.\color{black}
 %In the 2D temperature map divided by WVT binning algorism, we found a 7--8 keV high temperature region (hot region) with a width of ~700 kpc in the merger axis direction. We conclude that there are shock waves in the southeast and northwest since pseudo-pressure jumps are also seen at the edge of the hot region.
%私たちはWVT binning algorismで領域分けした2D温度マップにおいて衝突軸方向に700kpcの幅で広がった7--8 keVの高温な領域(hot領域)を発見した。高温領域の端でpseudo圧力もジャンプしていることから我々は南東と北西の衝撃波であると考えた。Channel

 \item We constructed the 3D original cluster model that reproduces the ICM distribution of the two clusters of galaxies before the collision, using the ICM profile in the outside region.
The ratio map of the 3D original cluster model to the observations showed that there is a clear enhancement of temperature and X-ray surface brightness in the hot region while those in the outside region is well reproduced by this simple 3D model.
%我々は元の密度勾配の影響を補正し、2つの衝撃波面で密度のサインを明確に表すため、3D cluster modelingと呼ぶ2つの衝突中の銀河団の元々のICM分布を構築した。表面輝度のTrend-diveマップでは、南東の銀河団はsloshingの兆候がある。sloshingの腕が伸びた先と南東の衝撃波面の交点には表面輝度のエンハンスが見られる。
 \item As the Mach numbers obtained by the temperature jump and density jump differ, the 3D merger model was constructed assuming that two temperature components, a shocked component and an unshocked component, overlap in the line of sight. 3D merger model showed that the shocked component with a temperature $kT_{\rm{s}}$ of $\sim$ 8 keV and a density $n_{\rm{s}}$ of $\sim 1.0\times10^{-3}~\rm{cm^{-3}}$ spreads to a depth of $\sim$ 1 Mpc on the shock wave front. We have shown that the SE and NW shock waves are estimated to have Mach numbers of $\sim$ 1.3 and $\sim$ 1.7, respectively.
%我々は視線上にpost-shockコンポーネントとpre-shockコンポーネントが重なっていると仮定した簡単な3D modelingを構築し、3D cluster modelingとスペクトルの2温度と合わせることでpost-shockコンポーネントの状態を推定した。3D modelingから温度が8keVで密度が1.0\times10^-3のpost-shockコンポーネントが1Mpcの奥行きをもつことを発見した。
 \item The $y$-map based on the 3D merger model agrees with the SZ signals observed by Plank, within the CMB fluctuation level.
%3D modelingにおいて衝撃波のマッハ数は南東で~1.3、北西で~1.7であった。post-shock領域を2つの衝撃波で挟み込んだモデルで計算されるymapはPlanckで観測されたymapとCMBゆらぎの範囲内で一致する。
 \item From the 3D merger model, the velocity of the shock wave was estimated to be $\sim1000~\rm{km~s^{-1}}$ for the SE and $\sim900~\rm{km~s^{-1}}$ for the NW, respectively, and the merging velocity was estimated to be $\sim1500~\rm{km~s^{-1}}$. From these velocities, the age of the shock wave is estimated to be $\sim$260~Myr \color{black} and the kinetic energy of the SE shock front is estimated to be $\sim$ $2.2\times10^{45} \rm{erg~s^{-1}}$. \color{black} If 10\% of this energy is used for a turbulence, the post shock turbulent velocity in the line of sight direction is calculated to be $\sim200~\rm{km~s^{-1}}$.
%衝撃波速度は南東で~1000km/s、北西で900km/sで衝突速度は~1500km/sであり、この速度より衝撃波速度は360 Myrと推定される。南東の衝撃波の持つ運動エネルギーは$2.2\times10^45 erg/sであり、この内10%が乱流に変換されたとすると視線方向の乱流速度は198km/sとなる。This value indicates that XRISM {\it resolve} observations of CIZA J1359 potentially be able to distinguish turbulent velocities in the post-shock and pre-shock region
\item The 3D merger model showed a local enhancement in some regions of the SE shock front (SE 6 and 7).
The ratio map of X-ray surface brightness using the 3D original cluster model shows a semi-clockwise spiral enhancement in the SE cluster, suggesting that there may have been local bulk motion in those regions prior to the collision.
\end{enumerate}
%CIZA1359の今後の観測期待

%・XRISM
%・GMRT$\cdot$MeerKATによる磁場観測(赤堀さんと藏原さん)

\begin{ack}
\normalsize{
This work is supported in part by JSPS KAKENHI Grant Numbers 15H03639, 15K17614, 21H01135, 16K05300, and 19K21054.}
\end{ack}

\appendix
\section*{Detail of 3D merger modeling}
\label{App:App1}
%\begin{figure}[ht]
 %\begin{center}
 %\includegraphics[width=8cm]{3Dmodeling_branch.jpg}
%\caption{}
 %\end{center}
%\caption{Appendix。}
%スペクトルの抽出領域をつける
%\label{.....}
%\end{figure}
%\begin{figure}[h]
% \begin{center}
% \includegraphics[width=8cm]{3Dmodeling_branch.png}
% \end{center}
% \caption{Conceptual diagram of 3D merger modeling.}
%\label{Appendix:3D modeling branch}

%\end{figure}

In this section, we describe the process flow of the 3D merger modeling outlined in section 3.2, using the results in the region SE 3 as an example. 
%The 3D merger modeling is outlined in section 3.2.

\begin{figure*}[t]
 %\begin{minipage}{0.5\hsize}
 %\begin{center}
 %\includegraphics[width=1\columnwidth]{3Dmodeling_branch.png}
  %\end{center}
%\end{minipage}
 %\begin{minipage}{0.33\hsize}
 \begin{center}
  \includegraphics[width=16cm]{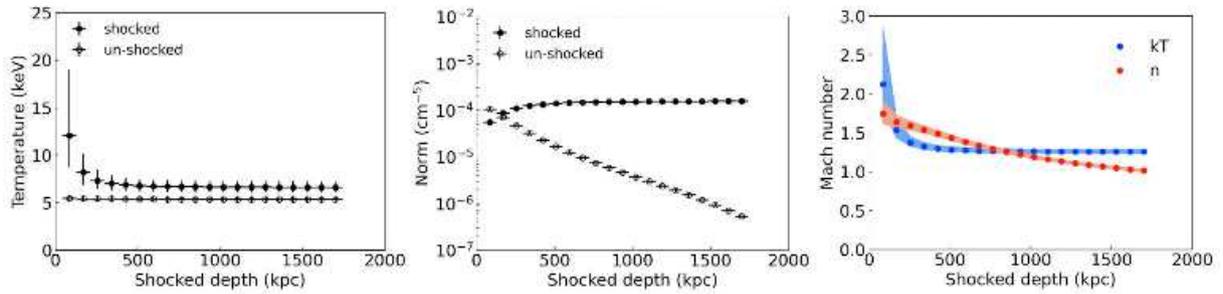}
 \end{center}
%\end{minipage}
%\begin{minipage}{0.33\hsize}
 %\begin{center}
 %\includegraphics[width=1\columnwidth]{appendix_8.png}
 %\end{center}
%\end{minipage}
%\begin{minipage}{0.33\hsize}
 %\begin{center}
 %\includegraphics[width=1\columnwidth]{appendix_.png}
 %\end{center}
%\end{minipage}
%\begin{minipage}{0.33\hsize}
% \begin{center}
% \includegraphics[width=1\columnwidth]{appendix_4.png}
% \end{center}
%\end{minipage}
%\begin{minipage}{0.33\hsize}
 %\begin{center}
 %\includegraphics[width=1\columnwidth]{appendix_5.png}
 %\end{center}
 %\end{minipage}
 \caption{Temperature (left) and norm (middle) distributions at each $l_{\rm{s}}$ obtained by a $2kT$ fitting with the unshocked model fixed in region SE 3. The unshocked component shows values estimated using the 3D cluster model. The Mach number distribution (right) is obtained from the R-H equation for temperature (kT) and density (n), respectively, at each depth using the 3D merger model.}
\label{Appendix:3D modeling}
\end{figure*}

\begin{figure}[t]
 \begin{center}
  \includegraphics[width=8cm]{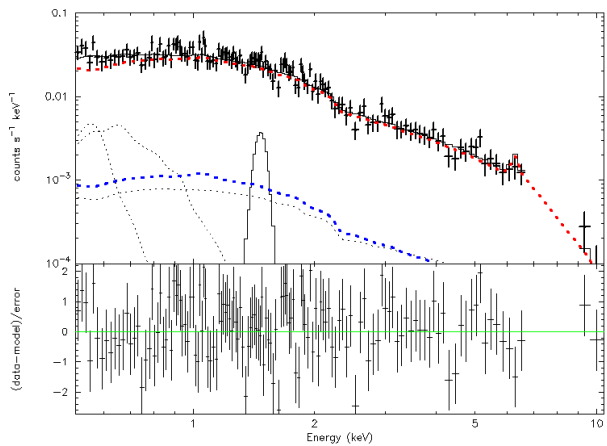}
 \end{center}
 \caption{The result of the spectrum fitting with the 2 $kT$ model in region SE 3, assuming $l_{s}$ is 10'. The red and blue dotted lines indicated shocked and unshocked models, respectively. Their parameters are listed in the upper column of table \ref{tab:first}}
%スペクトルを示す領域を付け加える
\label{spectrum_fit}
\end{figure}

\begin{table}[ht]
\tbl{Results of two temperatures fitting in the region 3
%\footnotemark[$*$] 
}{%
\scalebox{0.85}{
\begin{tabular}{lccccc}
\hline
\begin{tabular}{l} \multicolumn{1}{c}{$kT_{\rm s}$} \\ (keV) \end{tabular} & \begin{tabular}{l} \multicolumn{1}{c}{$kT_{\rm us}$} \\ (keV) \end{tabular} & \begin{tabular}{l} \multicolumn{1}{c}{$norm_{\rm s}$\footnotemark[$\dag$]} \\ \multicolumn{1}{c}{$10^{-5}$} \end{tabular} & \begin{tabular}{l} \multicolumn{1}{c}{$norm_{\rm us}$\footnotemark[$\dag$]} \\ \multicolumn{1}{c}{$10^{-5}$} \end{tabular} & \begin{tabular}{l} $\chi^{2}/\rm dof$ \end{tabular}\\
\hline
\begin{tabular}{l}
\multicolumn{1}{c}{$6.7^{+0.6}_{-0.6}$}\\ \end{tabular} &  5.3(fix)&$15.1^{+0.3}_{-0.3}$ & 0.58(fix) & 125/135~=~0.90\\
\multicolumn{1}{c}{$17.7^{+16.7}_{-9.4}$} & 5.3(fix) & $4.5^{+6.5}_{-1.9}$ & $11.6^{+2.2}_{-6.6}$ & 124/136~=~0.90\\
%\multicolumn{1}{c}{$17.7^{+16.7}_{-9.4}$}& 5.3(fix) & $4.5^{+6.5}_{-1.9}$ & $11.6^{+2.2}_{-6.6}$ & 124/136~=~0.90\\
\hline
\end{tabular}}}\label{tab:first}
\begin{tabnote}
%\footnotemark[$*$] This list is just a sample. \\
\footnotemark[$\dag$] Normalization of the {\it apec} thermal spectrum, which is given by $\lbrace10^{-14}/[4\pi(1+z)^{2}d_{A}^{2}]\rbrace\int{n_{\rm e}n_{\rm H}dV}$, where $d_{A}$ is the angular diameter distance, $n_{\rm H}$ is the ionized hydrogen density, V is the volume of the region.
\label{table:table3}
\end{tabnote}
\end{table}
According to item 1 in section 3.2, we estimated $norm_{\rm{us}}$ and $kT_{\rm{us}}$ at individual $l_{\rm s}$, scanning every 1' from 0' to 20' ($\sim$ 1600 kpc), as shown in open circle points in the left and middle panels of figure \ref{Appendix:3D modeling}.
%Following the bullet point 1 in section 3.2, we scanned $l_{\rm s}$ every 1' from 0'-20' ($\sim$ 1600 kpc) and estimated $norm_{\rm{us}}$ and $kT_{\rm{us}}$ at individual $l_{\rm{s}}$, as indicated by open circle points in the left and middle panels of figure \ref{Appendix:3D modeling}.
In this figure, the black circle points show the derived $norm_{\rm{s}}$ and $kT_{\rm{s}}$ by adjusting 2 $kT$ with $norm_{\rm{us}}$ and $kT_{\rm{us}}$ fixed at their best-fit values, at individual $l_{\rm{s}}$, according to item 2.
%The black circle points in the right and middle panels of figure \ref{Appendix:3D modeling} shows $norm_{\rm{s}}$ and $kT_{\rm{s}}$ at individual $l_{\rm{s}}$, performing 2 $kT$ fitting with $norm_{\rm{us}}$ and $kT_{\rm{us}}$ fixed at their best-fit values, according to item 2.
At $l_{\rm s}$ of 10', the results of the 2 $kT$ fitting are shown in figure \ref{spectrum_fit}, and the values of each parameter are listed in the upper column of table \ref{table:table3}.
%The results of the spectral fit assuming $l_{\rm s}$ is 10' are shown in figure \ref{spectrum_fit}, and the values of each parameter are listed in the upper column of table \ref{table:table3}.
%The results with $norm_{\rm{us}}$ free are shown in the lower column of table \ref{table:table3}. As shown these results, since $kT_{\rm s}$ is not well defined, we determined to fixed all parameters for the un-shocked model.
According to item 3, we derived two Mach numbers for the density and the temperature at each $l_{\rm s}$, as shown in the right panel of figure \ref{Appendix:3D modeling}.
We assumed that in the shocked condition, the temperature is uniformly spread over the emission-weighted temperature and the density has a gradient according to the 3D original cluster model.
Errors are evaluated by giving fitting parameters at three levels: maximum, minimum, and best-fit values for the 3D original cluster model, to the parameters of the unshocked regions and those of the preshock condition. We determined the intersection of the density and temperature Mach number curves as the shocked state parameters. However, there are two areas of intersection, one near 80--200 kpc and the other at $\sim$ 1 Mpc. If $l_{\rm{s}}$ is 80--200 kpc, the hot region must be thinner along the line of sight than along the merging axis. Since the 3D merger model assumes the major merger on the sky plane, we adopted the intersection near 1 Mpc as our parameter.

Here, can $kT_{\rm s}$ and $l_{\rm s}$ be determined by simply fitting the 2 $kT$ models with their parameters free? The lower column of table \ref{table:table3} shows the results of 2 $kT$ fitting with norms free in addition to the parameters of the shocked model.
The parameters of the shocked model, especially $kT_{\rm s}$, have a large error range. Therefore, we fixed both the $kT_{\rm us}$ and $norm_{\rm us}$, assuming the R-H relations.

We adapted these methods to the shock surface of the southeast (SE 1--9) and northwest (NW 1--5), respectively. 
As exemplified in table \ref{table:table3}, $\chi^{2}$/{\it d.o.f} value was reasonable in all cases, indicating a good fit. 
\color{black} 
These results are shown in figure \ref{result}.

\color{black}

\bibliographystyle{apj}
\bibliography{pasj}

\end{document}